\documentclass[useAMS,usenatbib,twocolumn]{mnras}
\usepackage{hyperref}
\usepackage{natbib}
\usepackage{graphicx}
\usepackage{booktabs}
\usepackage{amssymb}
\usepackage{caption}
\usepackage{times}
\usepackage{rotate}
\usepackage{url}
\usepackage{amsmath}
\usepackage{pbox}
\usepackage{cleveref}
\usepackage[dvipsnames]{xcolor}

{}
\newcommand{\overbar}[1]{\mkern 1.5mu\overline{\mkern-1.5mu#1\mkern-1.5mu}\mkern 1.5mu}

\newcommand{\redm}{redMaPPer}

\newcommand{\etal}{et al.}
\newcommand{\eg}{{\it e.g.,\/}}

\newcommand{\ie}{{\it i.e.,\/}}
\newcommand{\etc}{{\it etc.\/}}

\newcommand{\exsitu}{{\it ex situ\/}}
\newcommand{\insitu}{{\it in situ\/}}

\newcommand{\Msol}{\mbox{$M_{\odot}$}}

\newcommand{\MstarSchechter}{\ensuremath{M^{\ast}}}
\newcommand{\Mstar}{\ensuremath{M_{\ast}}}
\newcommand{\Mhalo}{\ensuremath{M_{\rm vir}}}

\newcommand{\M}[1]{\ensuremath{M_{\ast,\,\rm #1}}} 
\newcommand{\Mbar}[1]{\ensuremath{\overbar{M}_{\ast,\,\rm #1}}} 
\newcommand{\FullMhalo}{\ensuremath{\tlog{}(\Mhalo{} / \Msol{})}}
\newcommand{\FullM}[1]{\ensuremath{\tlog{}(\M{#1} / \Msol{})}}

\newcommand{\ngals}{\ensuremath{N_{\rm gals}}}

\newcommand{\tlog}{\text{log}}
\newcommand{\tlogten}{\text{log}_{10}}

\newcommand{\dex}{\text{dex}}
\newcommand{\mysim}{\ensuremath{{\sim}\,}}

\newcommand{\scatterMstarx}[1]{\ensuremath{\sigma_{\M{#1} | \Mhalo{}}}}

\newcommand{\scatterMstarcentitle}{\ensuremath{\sigma_{M_{\ast,\,\rm cen} | \Mhalo{}}}}

\newcommand{\scatterCenTot}{\ensuremath{\sigma_{\M{cen} | \M{tot}}}}

\newcommand{\deltaCenTot}{\ensuremath{\Delta_{\M{cen} | \M{tot}}}}
\newcommand{\deltaCenTotHP}{\ensuremath{\Delta_{\M{cen} | \M{tot},~\text{halo properties}}}}

\newcommand{\scatterMhaloLK}{\ensuremath{\sigma_{\Mhalo{} | L_K}}}

\newcommand{\scatterMhaloLambda}{\ensuremath{\sigma_{\Mhalo{} | \lambda}}}

\newcommand{\scatterMhalox}[1]{\ensuremath{\sigma_{\Mhalo{} | \M{#1}}}}

\newcommand{\Kpc}{\ensuremath{\mathrm{kpc}}}
\newcommand{\Mpc}{\ensuremath{\mathrm{Mpc}}}

\newcommand{\h}{\ensuremath{h^{-1}}}

\newcommand*{\fullref}[1]{\hyperref[{#1}]{\autoref*{#1} \nameref*{#1}}}

\hypersetup{draft}  
\begin{document}


\title[Physical Correlations of the Scatter between Galaxy Mass, Stellar Content, and Halo Mass]{Physical Correlations of the Scatter between Galaxy Mass, Stellar Content, and Halo Mass}

\author[Bradshaw, Leauthaud, Hearin, Huang, Behroozi]{Christopher Bradshaw$^{1}$\thanks{E-mail: christopher.bradshaw@ucsc.edu},
    Alexie Leauthaud$^{1}$, Andrew Hearin$^{2}$, Song Huang$^{1}$, Peter Behroozi$^{3}$
\\
\\
$^{1}$Department of Astronomy and Astrophysics, University of California, Santa Cruz, 1156 High Street, Santa Cruz, CA 95064 USA \\
$^{2}$High-Energy Physics Division, Argonne National Laboratory, Argonne, IL 60439, USA \\
$^{3}$Department of Astronomy and Steward Observatory, University of Arizona, Tucson, AZ 85721, USA \\
}

\maketitle\label{firstpage}


\begin{abstract}
\noindent We use the UniverseMachine to analyze the source of scatter between the central galaxy mass, the total stellar mass in the halo, and the dark matter halo mass. We also propose a new halo mass estimator, the {\it cen+N} mass: the sum of the stellar mass of the central and the $N$ most massive satellites.
We show that, when real space positions are perfectly known, the cen+N mass has scatter competitive with that of richness-based estimators.
However, in redshift space, the cen+N mass suffers less from projection effects in the UniverseMachine model.
The cen+N mass is therefore a viable low scatter halo mass estimator, and should be considered an important tool to constrain cosmology with upcoming spectroscopic data from DESI.
We analyze the scatter in stellar mass at fixed halo mass and show that the total stellar mass in a halo is uncorrelated with secondary halo properties, but that the central stellar mass is a function of both halo mass and halo age. This is because central galaxies in older halos have had more time to grow via accretion. If the UniverseMachine model is correct, accurate galaxy-halo modeling of mass selected samples therefore needs to consider halo age in addition to mass.
\end{abstract}

\begin{keywords}
    galaxies: clusters -- cosmology: observations -- large-scale structure
\end{keywords}







\section{Introduction}

The abundance of galaxy groups and clusters is a powerful tool for constraining cosmology, particularly the cosmological parameters $\sigma_{8}$ and $\Omega_{m}$ \citep[\eg{}][]{White1993, Rozo2009b, Weinberg2013}.
However, current cosmological constraints have uncertainties dominated by cluster mass uncertainties \citep[\eg{}][]{PlanckCollaboration2016a}.
To improve these constraints, the ideal halo mass estimator would have low intrinsic scatter in the observable -- \Mhalo{} relation and be easy to observe across a large fraction of the sky.

While scatter in the observable -- \Mhalo{} relation is a source of uncertainty in cosmology, it can also be an important source of information about galaxy formation and evolution.
For example, \citet{Tinker2017a} showed that measurements of the scatter in the \Mstar{} -- \Mhalo{} relation can be used to help constrain galaxy quenching, and \citet{Gu2016} showed that these observations, along with estimates of the scatter due to hierarchical assembly, can constrain the scatter in star formation.
Finally, correlations between scatter and halo or baryonic properties can suggest that the stellar content depends on properties other than the halo mass \citep[\eg{}][]{Croton2007, Zentner2014, Hoshino2015, Matthee2017, Kulier2018}.

Therefore, to better constrain cosmology and galaxy evolution, the development and analysis of accurate, large area halo mass estimators is an important area of research.
Estimators that use optical and near-IR data are of particular interest because of the wealth of both wide and deep data that will come from surveys such as
the Subaru Hyper Suprime-Cam Survey \citep[HSC,][]{Aihara2018},
the Dark Energy Survey \citep[DES,][]{Abbott2018},
the Large Synoptic Survey Telescope \citep[LSST,][]{Ivezic2019},
the Dark Energy Spectroscopic Instrument \citep[DESI,][]{DESI2016},
and \emph{Euclid} \citep{Laureijs2011}.
Optical and near-IR estimators can also probe a lower halo mass range than other methods such as X-rays \citep[\eg{}][]{Kravtsov2006, Mahdavi2013, Mantz2016}
and $Y_{SZ}$ \citep[\eg{}][]{Sunyaev1970, Marriage2011, Bleem2015}.
For these reasons, in this paper we focus on observables accessible to these next generation surveys.

The simplest optical proxy for halo mass is the stellar mass of the central galaxy (\M{cen}).
Previous work indicates that the lognormal scatter in \M{cen} at fixed halo mass (\scatterMstarx{cen}) is approximately $0.2~\dex{}$ \citep[\eg{}][]{More2009, Yang2009a, Behroozi2010, Guo2010, Leauthaud2012, Reddick2013}. In this paper, all estimates of scatter are in units of \dex{} and all logarithms are assumed to be base 10.
Expressing the scatter in \M{cen} at fixed halo mass is physically motivated because galaxy formation is known to exhibit strong dependence on halo mass \citep[\eg{}][]{White1978, Blumenthal1984}.
However, for a halo mass estimator the inverse of this is needed -- the scatter in halo mass at fixed \M{cen} (\scatterMhalox{cen}).
Assuming a power law relation $\M{cen} \propto \Mhalo{}^{\alpha}$ and $\alpha = 0.4$ from \citet{Kravtsov2018}, $\scatterMhalox{cen} \approx 0.5$ \dex{}.

A second optical proxy is cluster richness: a measure of the number of galaxies in a halo. One of the best current richness estimators is \redm{}'s $\lambda$ \citep{Rykoff2014}, which was tuned to minimize the scatter in halo mass.
Initial estimates in \citet{Rozo2014} and \citet{Rozo2015} using SDSS DR8 data found $\scatterMhaloLambda{} \approx 0.10$ for $\lambda > 20$ or $\Mhalo{} > 3 \times 10^{14} \Msol{}$.
However, more recent results from \citet{Mantz2016} and \citet{Murata2018} suggest $\scatterMhaloLambda{} \approx 0.2$ at this mass with a scatter that increases (decreases) at lower (higher) richness.

A third class of optical proxies uses the luminosity or mass of multiple members of the cluster.
The total stellar mass in the cluster (\M{tot}) was proposed by \citet{Andreon2012} and found by \citet{Kravtsov2018} to have $\scatterMhalox{tot} \approx 0.18$, significantly less than the scatter using \M{cen}.
\citet{Golden-Marx2018} showed that information from even a few satellites (parameterized by the magnitude gap) could also significantly reduce scatter in halo mass estimates.
An analogous measurement to the total stellar mass is the total K band luminosity from the cluster.
\citet{Ziparo2016} found that measuring luminosity within $1 \Mpc{}$ resulted in $\scatterMhaloLK{} = 0.18 \pm 0.07$ for $\Mhalo{} > 3.5 \times 10^{13} \Msol{}$.
At higher masses ($\Mhalo{} > 2 \times 10^{14} \Msol{}$), \citet{Mulroy2014} found a significantly lower scatter of $\scatterMhaloLK{} \approx 0.05$.

A summary of both optical and other halo mass estimators is shown in \autoref{tab:cluster_mass_summary}. For a detailed review of the performance of optical estimators and how they are impacted by projection, we refer readers to \citet{Pearson2015} and \citet{Wojtak2018}.

\begin{table*}
\begin{tabular}{l l l l}
    \toprule
    Observable                  & $\sigma_{\Mhalo{} | observable}$ [dex]& Halo Mass [$10^{13}$ \Msol{}] & Reference     \\ \midrule 

    $Y_{sz}$                    & $0.09 \pm 0.02$                       & $>20$          & \citet{Bleem2015}            \\ \midrule 

    $Y_X(<1\Mpc)$               & $0.11 \pm 0.05$                       & $>20$          & \citet{Mahdavi2013}           \\ \midrule 
    $L_X^{cut}(<1\Mpc)$         & $0.14 \pm 0.02$                       & $>20$          & \citet{Mahdavi2013}           \\ \midrule 

    $\M{tot}$                   & $0.18 \pm 0.06$                       & $>5$           & \citet{Kravtsov2018}          \\ \midrule 
    $L_K(<1\Mpc)$               & $0.18 \pm 0.07$                       & $>3.5$         & \citet{Ziparo2016}            \\ \midrule 
    $\lambda$                   & $0.11 \pm 0.02$                       & $>10$          & \citet{Rozo2014}    			\\ \midrule
    $\lambda$                   & $0.2 \pm 0.02$                        & $30$           & \citet{Murata2018}    	    \\ \midrule

    $\M{cen}$                   & $0.45 \pm 0.13$                       & $>5$           & \citet{Kravtsov2018}          \\ \bottomrule 
\end{tabular}
\caption{
    Scatter in the halo mass estimates of various proxies.
    Estimates using the Sunyaev-Zel'dovich effect and X-ray properties have extremely low scatter $\approx 0.1$ \dex{} but are limited to high mass clusters and are not available to optical surveys.
    Estimates with optical proxies that use information from multiple galaxies in the halo (\M{tot}, $L_K$, $\lambda$) have scatter $\approx 0.18$ (using the more conservative \citet{Mantz2016} and \citet{Murata2018} estimates for $\lambda$).
    While it is a simple observable, estimates using \M{cen} have a much larger scatter than the other proxies.
\label{tab:cluster_mass_summary}
}
\end{table*}

While many studies have characterized the amount of scatter between various observables and halo mass, the source of the scatter is much less understood.
An improvement in our understanding of the factors that cause the scatter would be valuable as it could directly improve our understanding of galaxy physics, and indirectly allow us to construct better halo mass estimators for cosmology. Perhaps the best-studied source of scatter is the distribution of secondary halo properties, such as age and concentration, among galaxies of the same \Mhalo{} \citep[\eg{}][]{Croton2007, Zentner2014, Hearin2016, Matthee2017}.
More recently, with large and accurate hydrodynamical simulations, baryonic properties are also being investigated as a source of the scatter \citep[\eg{}][]{Kulier2018}.
Scatter also naturally arises from intrinsic stochasticity in hierarchical assembly \citep[\eg{}][]{Gu2016} and galaxy quenching \citep[\eg{}][]{Tinker2017a}.

In this paper, we both propose a new stellar-mass-based halo mass estimator, and investigate the contribution that variance in secondary halo properties makes to the scatter in stellar mass (both \M{cen} and \M{tot}).
The new estimator is the {\it cen+N} mass
(\M{N}), defined as the sum of the mass of the central and the $N$ most massive satellites. We show that, with only a few satellites, this is a competitive halo mass proxy.
We then show that \M{N} and \M{tot} have significantly less dependence on secondary halo properties than \M{cen}.
Using these findings, we further show that the scatter in \M{cen} can be decomposed into a stochastic component (due to hierarchical assembly) and an age dependent process (related to the mergers of satellite galaxies onto the central).

This paper is organized as follows.
In \autoref{sec:simulations} we introduce the UniverseMachine simulation on which our analysis is based.
In \autoref{sec:results} we present and analyze the performance of the cen+N mass proxy.
In \autoref{sec:biases} we show that both \M{N} and \M{tot} are less sensitive to secondary halo properties than \M{cen}, and in \autoref{sec:origin} we use this to decompose the scatter into a stochastic and age dependent process.
Finally, we summarize and conclude in \autoref{sec:conclusions}.

We adopt a flat $\Lambda{}$CDM, {\it Planck\/} cosmology ($\Omega_M = 0.307, \Omega_B = 0.048, \Omega_\Lambda = 0.693, \sigma_8 = 0.829, n_s = 0.96, H_0 = 67.8$) \citep{PlanckCollaboration2016}.

\section{Simulations}\label{sec:simulations}

\subsection{Small MultiDark Planck (SMDPL)}\label{subsec:smdpl}

The SMDPL\footnote{doi:10.17876/cosmosim/smdpl/} \citep{Klypin2016, Rodriguez-Puebla2016} N-body simulation contains $3840^3$ (${\sim}\,6 \times 10^{10}$) particles in a periodic, comoving volume $400~\h{}~\Mpc{}$ on a side.
It was run with the GADGET-2 code \citep{Springel2005a} and has excellent mass ($9.6 \times 10^7 \Msol$) and force ($1.5~\h{}~\Kpc$) resolution. SMDPL uses a cosmology consistent with \citet{PlanckCollaboration2016}.

Halos were found using Rockstar and merger trees constructed with Consistent Trees \citep{Behroozi2013b, Behroozi2013c}.
We use a snapshot at $z \approx 0.40$ to match the HSC analysis of \citet{Huang2018}.

We convert the default Rockstar mass accretion rate to a unitless measurement as in \citet{Diemer2017a}:

\begin{equation}
\label{eq:gamma_definition}
\Gamma = \frac{\tlog[M(t)] - \tlog[M(t - \Delta t)]}{\tlog[a(t)] - \tlog[a(t - \Delta t)]} \\
\end{equation}

\noindent where $M(t - \Delta t) = M(t) - \Gamma_{\rm{Rockstar}, \Delta t} \cdot \Delta t$. Unless specified, we use $\Delta t = 2\, t_{\rm dyn, rockstar}$ where

\begin{equation}
    t_{\rm dyn} = (\frac{4}{3} \pi{} G \Delta_{\rm c} \rho_{\rm m})^{-\frac{1}{2}}
\end{equation}

\noindent and $\Delta_{\rm c}$ is the virial overdensity in units of $\rho_{\rm crit}$ using the \citet{Bryan1998} fitting formula and $\rho_{\rm m} = \rho_{\rm crit} \Omega_{\rm m} {(1 + z)}^3$ is the mean matter density. For an overview of the variety of dynamical times used throughout the literature see Xhakaj \etal{} (in prep.) \footnote{$2\, t_{\rm dyn, rockstar}$ corresponds to a crossing time or $1\, t_{\rm dyn, COLOSSUS}$ \citep{Diemer2018}}.

\subsection{The UniverseMachine}\label{subsec:um}

The UniverseMachine\footnote{https://www.peterbehroozi.com/data.html} \citep[UM;][]{Behroozi2018} is an empirical model that predicts the star-formation histories of galaxies across cosmic time. The foundation of the UM is a flexible, parameterized model for the connection between galaxy star formation rates (SFR) and the assembly history of dark matter halos. In this model, SFR is a function of $v_{\text{Mpeak}} \equiv v_{\text{max}}(z_{\text{Mpeak}})$, the maximum circular velocity at the redshift when the halo attained its peak mass, $\Delta v_{\text{max}}$, the growth in $v_{\text{max}}$ in the last dynamical time, and redshift. With a functional form for $SFR(v_{\text{Mpeak}}, z, \Delta v_{\text{max}})$, UM maps an \insitu{} SFR to each halo and subhalo at each snapshot of the simulation. At any given snapshot, the stellar mass of a galaxy is calculated by integrating the star-formation history of the galaxy, additionally accounting for \exsitu{} mass growth from mergers, and mass loss from passive evolution \citep[see][for further details]{Behroozi2018}.

The parameters of the UM model were fit to a diverse set of observations from a data compilation spanning a wide range of redshifts, $0 < z\lesssim10$. These data include stellar mass functions from ZFOURGE/CANDELS \citep{Tomczak2014} and PRIMUS \citep{Moustakas2013}; quenched fractions in PRIMUS \citep{Moustakas2013} and COSMOS/UltraVISTA \citep{Muzzin2013}; and galaxy correlation functions from SDSS DR7 \citep{Abazajian2009}.

The UM output that we use differs slightly from that discussed in \citet{Behroozi2018} in that it separately tracks the \insitu{} and \exsitu{} growth, rather than placing some fraction of the \exsitu{} mass in the central galaxy and the rest in the intracluster light.
This model overestimates the number density of very high mass galaxies compared to, for example, the HSC survey\footnote{See \citet{Huang2018} for a rescaling of the UM masses to fit HSC. We do not use these rescaled masses here.}. This may partly be due to a difference in the mass definition: the UM includes all stellar mass while HSC will miss some of the diffuse component.
While this steeper slope of the \M{cen} -- \Mhalo{} relation will likely lead to a lower absolute \scatterMhalox{cen}, we primarily focus on relative comparisons which remain valid.

\section{Comparison of Halo Mass Proxies}\label{sec:results}

One of the primary goals of this paper is to investigate new, low scatter, halo mass proxies. In this section we present our candidate, the {\it cen+N mass}, which is defined as the sum of the mass of the central and the $N$ most massive satellites.
We motivate this choice by demonstrating that stellar-mass-based proxies that include more of the stellar mass in the halo have reduced intrinsic scatter, and argue that the cen+N mass makes the right trade-off in reducing the intrinsic scatter while keeping the cluster finding requirements simple enough to minimize projection effects.
Finally, we compare the scatter, both intrinsic and with projection effects, of the cen+N mass to that of a richness-based estimator.

\subsection{Notation for Masses and Scatter}

We define \insitu{} stellar mass as stars that formed in the central galaxy of the host halo and \exsitu{} stellar mass as stars deposited onto the central galaxy by mergers \citep{Huang2018, Rodriguez-Gomez2016}. We use the following notation throughout:

\begin{itemize}
    \item \M{in}: the \insitu{} stellar mass of the central galaxy.
    \item \M{ex}: the \exsitu{} stellar mass of the central galaxy.
    \item \M{cen}: the stellar mass of the central galaxy $= \M{in} + \M{ex}$
    \item \M{sat}: the sum of the stellar mass of all satellite galaxies in the halo.
    \item \M{tot}: the total stellar mass in the halo $= \M{cen} + \M{sat}$
    \item \M{N}: the cen+N mass, the sum of \M{cen} and the stellar mass of the $N$ most massive satellites.
    \item \M{x}: A generic stellar mass (any of \M{in}, \M{cen}, \M{2}, \etc{})
    \item $\sigma_{x | y}$: the lognormal scatter, in dex, of $x$ at fixed $y$.
\end{itemize}

\subsection{Central and Total Stellar Mass}\label{subsec:totalmstar}

\begin{figure*}
    \begin{center}
    \includegraphics[width=\textwidth]{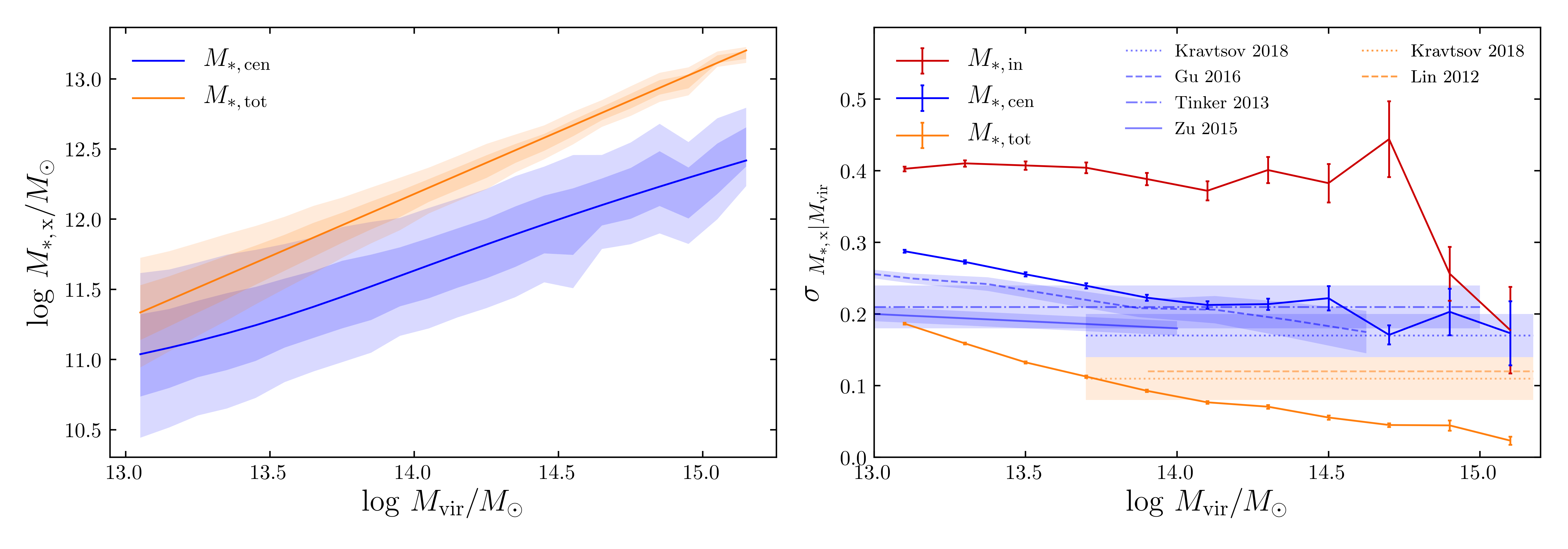}
    \caption{\
        Comparison of the slope and scatter of the total, central and \insitu{} stellar mass as a function of \Mhalo{} in the UM and literature.
        \textbf{Left}:\
        In group and cluster mass halos, \M{tot} increases more steeply with halo mass, and has less scatter at fixed halo mass, than \M{cen}. The solid lines and shaded regions show the best fit using the \citet{Behroozi2010} functional form, and the one and two sigma scatter.
        \textbf{Right}:\
        \scatterMstarx{in} is significantly larger than \scatterMstarx{cen}, which implies that the observed scatter in the \M{cen} -- \Mhalo{} relation may depend on observation depth. Shallow observations that only capture the bright inner portion of the central will find a larger scatter than those that see more of the light.
        \M{tot} is most tightly correlated with \Mhalo{} and is an excellent tracer of halo mass with scatter ranging from 0.19 to 0.04 dex at $\FullMhalo{} = 13$ and $15$ respectively.
        \scatterMstarx{cen} and \scatterMstarx{tot} in the UM are broadly consistent with that found in the literature. However, while most previous works assumed mass independent scatter, the UM predicts a significant decrease in scatter with increasing \Mhalo{}.
    }
    \label{fig:intro_plot}
    \end{center}
\end{figure*}

We first characterize the \M{cen} -- \Mhalo{} and \M{tot} -- \Mhalo{} relations in the UM\@.
We then test the performance of estimators based on \M{cen} and \M{tot}. We show that the UM has scatter consistent with results in the literature.

We find the best fit to the \M{x} -- \Mhalo{} relation for each mass definition (\eg{} \M{cen}, \M{2}) for the five parameter functional form from \citet{Behroozi2010} and widely used in the literature \citep[\eg{}][]{Leauthaud2011, Geha2012}:

\begin{equation}
\label{eq:sm_hm_functional_form}
\tlogten{}(\Mhalo{}) = \tlogten{}(M_1) + \beta~\tlogten{}(\frac{\M{x}}{\M{0}}) + \frac{(\frac{\M{x}}{\M{0}})^{\delta}}{1 + (\frac{\M{x}}{\M{0}})^{- \gamma}} - \frac{1}{2}
\end{equation}

\noindent where $M_1$ is a characteristic halo mass, \M{0} a characteristic stellar mass, $\beta$ controls the low mass slope, $\delta$ the high mass slope, $\gamma$ the transition from the low to high mass regime, and $\M{x}$ is the stellar mass under consideration.

The best fits for \M{cen} and \M{tot}, along with the one and two sigma scatter, are shown in the left panel of \autoref{fig:intro_plot}. We find that these fits differ in a few ways.
First, a power law (a special case of \autoref{eq:sm_hm_functional_form} with $\delta = \gamma = 0$) is sufficient for the \M{tot} -- \Mhalo{} relation, but all five parameters are required for \M{cen}.
Second, \M{tot} increases more steeply with \Mhalo{} than \M{cen};
$d\log{}\M{tot}/d\log{}\Mhalo{} \approx 0.89$ at all halo masses, while $d\log{}\M{cen}/d\log{}\Mhalo{}$ varies between $0.5$ at $\FullMhalo{} \approx 13$ and $0.75$ at $\FullMhalo{} \approx 14$.
Third, \scatterMstarx{cen} is significantly larger than \scatterMstarx{tot}.

The right panel of \autoref{fig:intro_plot} compares the scatter of the \insitu{}, central, and total stellar mass estimators. This shows more clearly that \scatterMstarx{cen} is roughly 0.1 \dex{} larger than \scatterMstarx{tot} at all halo masses. However, this figure also shows that, for both estimators, the scatter decreases significantly as halo mass increases: from 0.28 to 0.18 \dex{} for \M{cen} and 0.19 to 0.04 \dex{} for \M{tot} in the mass range $\FullMhalo{} = 13$ to 15.

The right panel also shows that \scatterMstarx{in} is large compared to \scatterMstarx{cen}. This is a potential problem for surveys that only detect the bright inner region of galaxies where a significant fraction of stellar mass comes from the \insitu{} component \citep[][]{Rodriguez-Gomez2016}.
Halo mass predictions using shallow surveys may therefore have significantly higher scatter than those using deeper surveys.
Huang \etal{} (in prep) found this effect in HSC observations: the scatter of the stellar mass within the inner 10 \Kpc{} is $\sim 30\%$ larger than the scatter using the maximum radius of the central galaxy.
Even in deep surveys, the exact results will be sensitive to the amount of light that is counted as part of the central galaxy vs the intracluster light (ICL). The output of the UM that we use includes all stellar mass that has merged with the central galaxy. In practice, some of this mass in the ICL will not be directly observed and is either ignored, or needs to be fitted for (Ardila \etal{} in prep).

Despite these concerns, the UM results are broadly consistent with previously published values. $\scatterMstarx{cen} \approx 0.2$ has been found using a variety of techniques \citep[\eg{}][]{Tinker2013, Zu2015, Gu2016, Kravtsov2018}, and $\scatterMstarx{tot} \approx 0.1$ was found by both \citet{Lin2012} and \citet{Kravtsov2018}. At $\FullMhalo{} \approx 14$ the scatter we measure is comparable to these fiducial values.

Our results differ from the literature in that, while many previous works assume mass independent scatter, we find a strong mass dependence. We discuss potential physical reasons for this decreasing scatter in \autoref{sec:origin}.

These results suggest that estimators that use more of the stellar mass (assuming perfect knowledge of cluster membership) display reduced scatter. In the next section we evaluate how quickly cen+N mass based estimators converge to the performance of \M{tot}. We also consider how these estimators degrade with uncertain redshifts and imperfect cluster finding.

\subsection{Cen+N Stellar Mass}\label{subsec:partial_stellar_content}

\begin{figure*}
    \begin{center}
    \includegraphics[width=\textwidth]{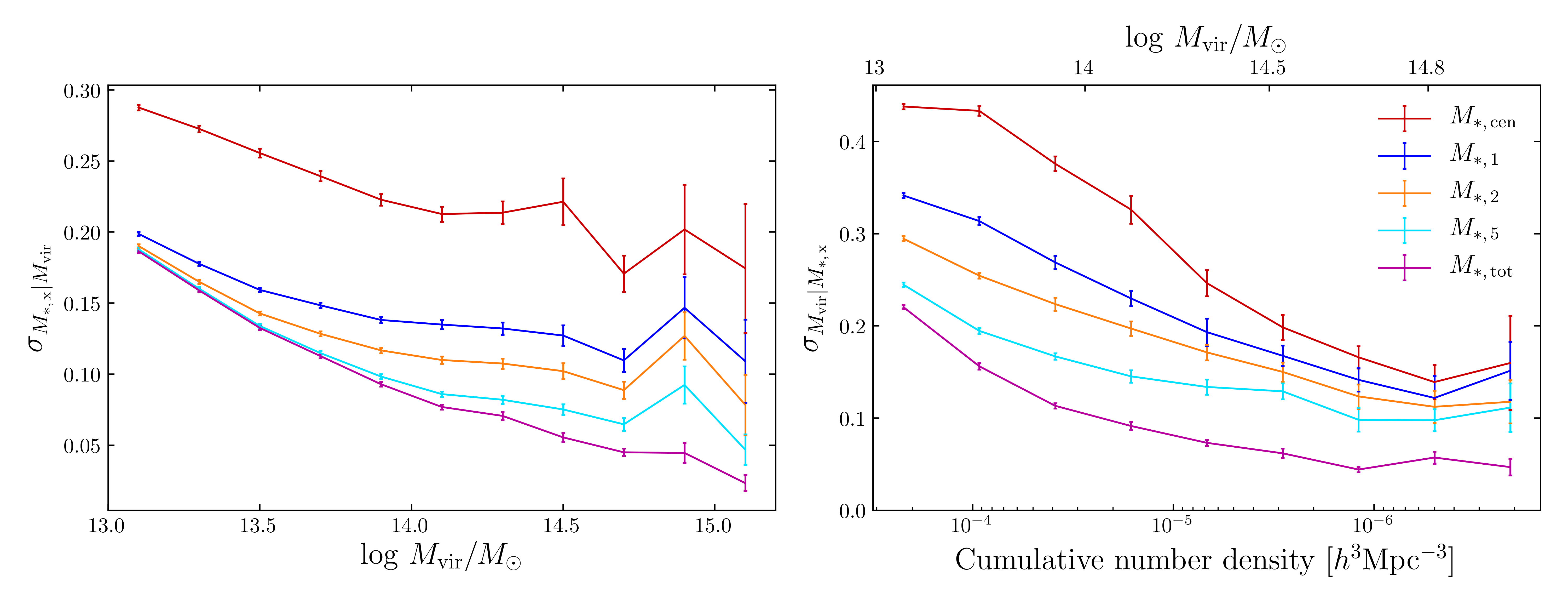}
    \caption{\
        The scatter in \M{x} at fixed \Mhalo{} (and vice-versa) for \M{cen}, selected definitions of the cen+N mass ($N = 1, 2, 5$), and \M{tot}.
        \textbf{Left}: At fixed \Mhalo{}, definitions of stellar mass that include more of the total stellar content of the halo have less scatter (\eg{} $\scatterMstarx{tot} < \scatterMstarx{2} < \scatterMstarx{cen}$).
        At $\FullMhalo{} \approx 13$, the cen+2 stellar mass has comparable scatter to \M{tot}.
        As halo mass increases, more satellites are needed to converge to the scatter of \M{tot}, though \scatterMstarx{5} is within $0.05$ dex of \scatterMstarx{tot} at $\FullMhalo{} \approx 15$.
        \textbf{Right}: At fixed cumulative number density (calculated from the stellar mass) using a stellar mass proxy that includes more of the stellar mass reduces the scatter in \Mhalo{}.
        However, the scatter in \Mhalo{} for the cen+N mass proxies converges slower than the scatter in the proxy at fixed \Mhalo{} shown on the left. This is primarily due to the increasing slope of of the \M{x} -- \Mhalo{} relation for proxies that contain more of the stellar mass.
        \label{fig:scatter_various_sats}
        }
    \end{center}
\end{figure*}

In the previous section, we showed that there is significantly less intrinsic scatter in \M{tot} at fixed \Mhalo{} than in \M{cen}.
However, determining \M{tot} in observations requires assigning cluster memberships to galaxies.
This process can introduce its own, potentially hard to quantify, uncertainties and biases (see \citealt{Wojtak2018} for a discussion of how imperfect cluster membership affects mass estimates).
Stellar mass based estimators therefore need to make a trade-off between reducing the intrinsic scatter of the observable (by including satellite masses) and adding scatter and bias in the cluster finder.
A potential compromise is the cen+N mass: the sum of \M{cen} and the $N$ most massive satellites.
For small $N$ this observable should be significantly less prone to cluster membership errors than \M{tot}.
This is particularly true with upcoming large spectroscopic surveys, such as DESI, which will obtain spectroscopic redshifts for the brightest cluster members (subject to observational constraints such as fiber collisions, the effects of which we do not model in this paper).
In this section we show that this observable has a competitive intrinsic scatter, even with relatively small $N$.

The left panel of \autoref{fig:scatter_various_sats} shows \scatterMstarx{x} for the central, total, and cen+N ($N = 1, 2 \textrm{ and } 5$) stellar masses.
The cen+N mass estimators have scatter between that of \M{cen} and \M{tot} with scatter decreasing as $N$ increases (\ie{} as the estimator includes a larger fraction of \M{tot}).
However, the importance of the $N$th satellite is not constant with halo mass.
At $\FullMhalo \approx 13$, including a single satellite reduces scatter to within 0.02 dex of \M{tot}.
In more massive halos, where a larger fraction of the total mass is outside the central, more satellites need to be added to converge to \M{tot}: at $\FullMhalo \approx 14$, \scatterMstarx{5} is within 0.01 dex of that of \M{tot}.
However, even in these massive halos, adding just a single satellite gives a significant improvement ($\sim{} 0.1~\dex{}$) over \M{cen}.

While the left panel shows the physically motivated \scatterMstarx{x}, the relevant quantity to evaluate a halo mass estimator is the scatter in \Mhalo{} at fixed observable (\scatterMhalox{x}) shown in the right panel. As the dependent variable is not consistent, this is plotted against number density, with the corresponding \Mhalo{} shown along the top axis to allow comparison to the left panel.

The relationship between the left and right panels is not obvious. For example, \scatterMstarx{1} is comparable to \scatterMstarx{tot} at $\FullMhalo = 13$ (left panel), but \scatterMhalox{1} is significantly different to \scatterMhalox{tot} at the same halo mass (right panel). This can be explained by assuming that, locally, the \M{x} -- \Mhalo{} relation is a power law,

\begin{equation}
    \tlogten{}(\Mstar{}) \propto \beta~\tlogten{}(\Mhalo{})
\end{equation}

\noindent with lognormal scatter. We can then convert between \scatterMstarx{x} and \scatterMhalox{x} as follows:

\begin{equation}
    \scatterMhalox{x} = \scatterMstarx{x} \frac{d\ \tlogten(\Mhalo)}{d\ \tlogten(\M{x})} =
    \frac{\scatterMstarx{x}}{\beta}.
\end{equation}
The quantity \scatterMhalox{x} is therefore a function both of \scatterMstarx{x}, and the slope, $\beta$.

As shown in \autoref{fig:intro_plot}, the slope steepens as more stellar mass is included. So, even though we have $\scatterMstarx{tot} \approx \scatterMstarx{1}$, we nonetheless have

\begin{equation}
    \scatterMhalox{tot} \approx \frac{\beta_{1}}{\beta_{\rm{tot}}} \scatterMhalox{1} < \scatterMhalox{1}
\end{equation}

\noindent as $\beta_{1} < \beta_{\rm tot}$.

Because of the slope-dependence of the scatter in halo mass, the quantity \scatterMhalox{x} does not converge as quickly to the performance of the total stellar mass as \scatterMstarx{x} does.
However, the cen+N mass is still a significant improvement over \M{cen}, and can be comparable to the low scatter attained with \M{tot}.
For example, at $\FullMhalo{} = 14$, $\scatterMhalox{5} \approx 0.17$, which is roughly half the scatter in halo mass at fixed \M{cen}, and only 0.06 dex larger than the scatter at fixed  \M{tot}.
Thus, even with relatively small $N$, the cen+N mass gives low scatter estimates of \Mhalo{}.

\subsection{Cen+N Versus Richness}\label{subsec:partial_stellar_content_vs_richness}

\begin{figure*}
    \begin{center}
    \includegraphics[width=\textwidth]{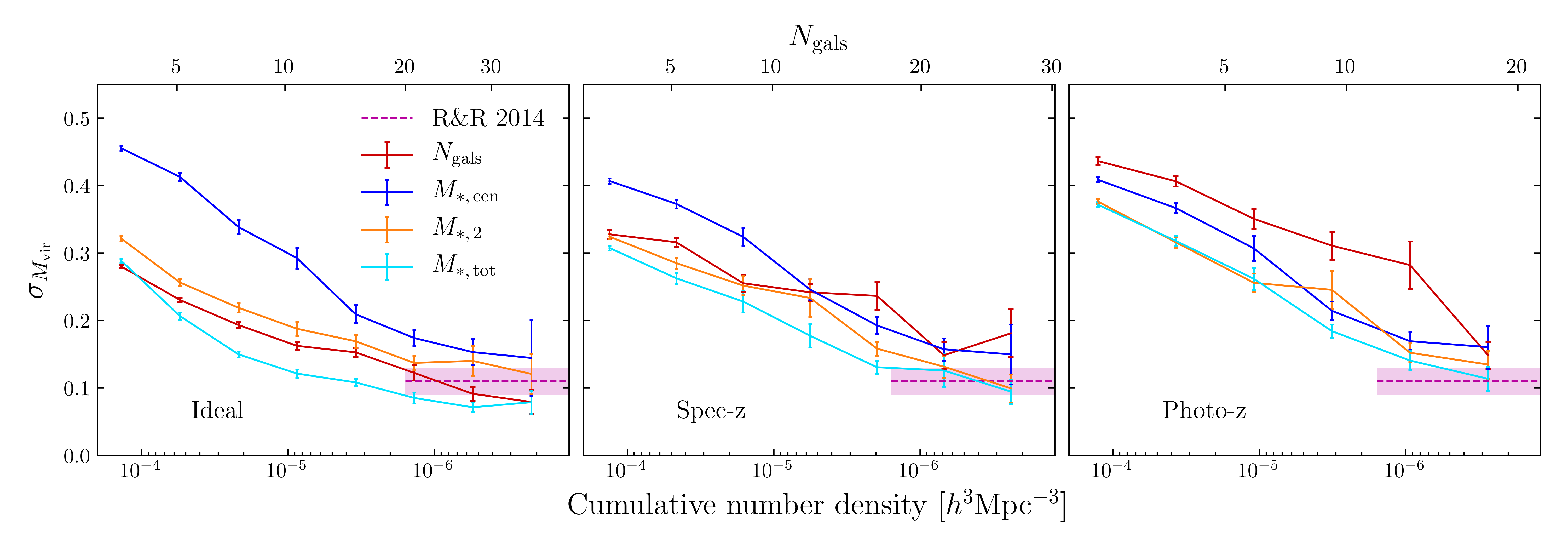}
    \caption{Scatter in \Mhalo{} at a fixed value for an observable (\M{x} or \ngals{}).
        \textbf{Left}: The ideal case which uses the real space positions and true cluster membership from the simulation. \M{tot} has the lowest intrinsic scatter while \M{2} is $\sim 0.03$ \dex{} worse than \ngals{} at all but the highest masses.
        \textbf{Center}: The case in which galaxies have spectroscopic redshifts.
        \M{2} suffers less from projection effects than, and is now an improvement on, \ngals{}.
        \textbf{Right}: A Simulation of a photometric survey where the redshift measurement includes an uncertainty of 90\h{}\Mpc{}. Because of the simplified cluster finding, none of the estimators are a significant improvement over \M{cen}. See the text for details of the cluster finder and the richness proxy \ngals{}.
        Shown for comparison is the scatter for $\lambda$ from \citet{Rozo2014} estimated from SDSS DR8 data. However, we caution that this scatter was calculated from generally much larger X-ray and SZ clusters (20 $< \lambda <$ 100) than those shown here.
        \label{fig:richness_comparison}
    }
    \end{center}
\end{figure*}

The current most popular halo mass proxy for large optical surveys is cluster richness. In particular, the red-sequence based \redm{} \citep{Rykoff2014} has been used with success on a number of surveys \citep[\eg{}][]{Rozo2014, Rykoff2016, McClintock2019}.
In this section, we compare the cen+N mass to a simple richness-based estimator. We include tests with simple models of projection effects.

We define our richness proxy, \ngals{}, as the number of galaxies that are more massive than some cutoff, and that have a specific star formation rate (sSFR) below some cutoff. We choose the same mass cutoff as in \redm{}:
$0.2 \MstarSchechter{} = 10^{10.64}$, 
and a cutoff in sSFR of $10^{-11}$ to ensure we only select red galaxies.
We apply this same mass cut (simulating survey completeness) to the galaxies included in the cen+N mass, though in practice this only affects the \M{tot} mass.
We emphasize that \ngals{} is not designed to precisely mimic \redm{}'s $\lambda$. To do this we would need to assign cluster membership using \redm{}'s algorithm, which is not possible as the UM (and no currently available model) can predict realistic galaxy colors.
Instead, we use a simple cluster finder (selection within some volume around known centrals) and compare the halo mass estimates of the generic richness estimator \ngals{} to those of the cen+N mass.

We test the performance of these estimators in three observing conditions.
First, we use the real space positions and true cluster memberships as given by the UM\@. This is an ideal case, unachievable in observations, but will show the intrinsic scatter for these estimators.
Second, we simulate a spectroscopic survey with precise redshift measurements. Observed real space positions are therefore only affected by redshift space distortions (RSD) \citep{Kaiser1987}.
Third, we simulate a photometric survey with redshift uncertainty of $\sigma_{z} / (1+z) = 0.01$. At our $z \approx 0.4$, this corresponds to $\approx$ 90\ \h{}\ \Mpc{} which is the estimated uncertainty of a single SDSS red galaxy at the median redshift of the \redm{} cluster sample given in \citet{Busch2017}.

In both the spectroscopic and photometric cases, we assume a cluster finder that can perfectly identify centrals. All galaxies within a cylinder centered on the central are considered members of this cluster. We use the virial radius (which in practice could be approximated iteratively) as the radius of this cylinder and 10\ \h{}\ \Mpc{} and 50\ \h{}\ \Mpc{} as the half-length for the spectroscopic and photometric surveys respectively.
While these cylinder choices have some impact on our results (\eg{} in the spectroscopic case, a shorter cylinder giver better results for lower mass clusters) we are not overly sensitive to changes in the ranges 5 -- 15\ \h{}\ \Mpc{} (spectroscopic) and 40 -- 90\ \h{}\ \Mpc{} (photometric).

The results for the ideal case with known 3d positions and cluster membership are shown in the left panel of \autoref{fig:richness_comparison}. Among the estimators considered, \M{tot} has the lowest scatter and is $\sim{} 0.05~\dex{}$ better than \ngals{} averaged across the mass range.
\M{2} has slightly worse performance than \ngals{}, within $0.03~\dex{}$ at all but the highest masses.
At intermediate richness ($5 < \ngals{} < 20$), the cen+N mass with between 2 and 5 satellites has intrinsic scatter comparable to that of richness; at higher richness, more satellites are needed, but these halos also have more bright satellites.

The middle panel of \autoref{fig:richness_comparison} compares the performance of the estimators assuming a spectroscopic survey with the effects of RSD and imperfectly assigned cluster membership.
As expected, all estimators (except \M{cen} which suffers no projection effects) show increased scatter compared to the ideal case. However, the performance of the \ngals{} estimator is now worse than \M{2} at all masses.
However, while we expect to have spectroscopic redshifts for the handful of members used in the cen+N mass estimate for DESI, a richness-based estimator needs redshifts for many more galaxies. Existing richness based catalogs usually rely on color, and so their performance will be a combination of the central and right panel which shows the performance with this larger uncertainty on position.

We note that \citet{Rykoff2012} mentioned that weighting cluster members by luminosity increased scatter -- the opposite to what we find. This could be due to a number of reasons:
1. Our simple cluster finding algorithm could be biased in favor of the cen+N mass method.
2. The SDSS photometry used in \citet{Rykoff2012} is missing light from low surface brightness outer regions \citep{Bernardi2013, Kravtsov2018} which would reduce the benefits of luminosity weighting since \autoref{fig:intro_plot} shows that \M{in} has a larger scatter.
3. Luminosity (or mass) weighting is likely more valuable at the lower mass range we are testing (as mentioned in \citealt{Rykoff2012}).
4. A large scatter in the mass -- luminosity relation decreases the value of mass weighting in observations.

We have shown that in the UM, and in the case of an idealized cluster-finder with zero projection effects, the cen+N mass, with N between 2 and 5, has intrinsic scatter comparable with \ngals{} at all but the highest halo masses ($\ngals{} > 20$, $\FullMhalo{} > 14.6$).
Moreover, when using a simple model for projection effects to relax the assumption of perfect cluster membership assignment, we find that the cen+N mass estimator can outperform richness-based mass estimation.
These results, combined with the fact that stellar mass is easier to model than richness as it does not rely on color, make the cen+N mass a promising halo mass estimator.

\section{Correlations Between Stellar Mass and Secondary Halo Properties}\label{sec:biases}

\begin{figure*}
    \begin{center}
    \includegraphics[width=\textwidth]{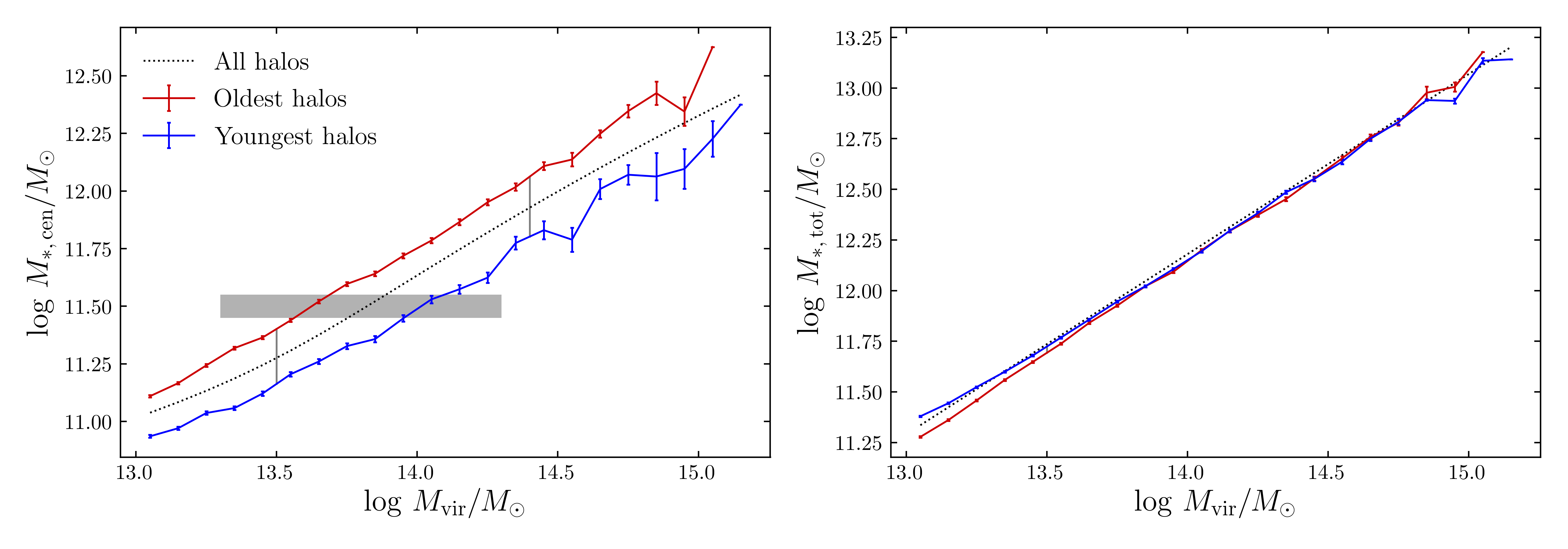}
    \caption{\
        The \M{cen} -- \Mhalo{} (left) and \M{tot} -- \Mhalo{} (right) relations for the youngest and oldest 20\% of halos in each mass bin. \M{cen} has a strong dependence on halo formation time; at all halo masses the oldest halos have a central stellar mass $\mysim{} 0.2~\dex{}$ greater than the youngest halos. In contrast, \M{tot} has almost no correlation with halo age.
        The vertical gray lines show the effect size of halo age at $\FullMhalo{} = 13.5~\text{and}~14.4$. \autoref{fig:effect_size} shows how the effect size varies for different secondary properties and stellar mass definitions.
        The horizontal gray box shows that an \M{cen} cut at $\FullM{cen} \approx 11.5$ selects old halos at a significantly lower mass than it does young halos. Because of this, and the shape of the halo mass function, \M{cen} selects an sample of halos that is biased old compared to a random selection of the same \Mhalo{} distribution.
        \label{fig:age_split_smhm}
    }
    \end{center}
\end{figure*}

\begin{figure}
    \begin{center}
    \includegraphics[width=0.5\textwidth]{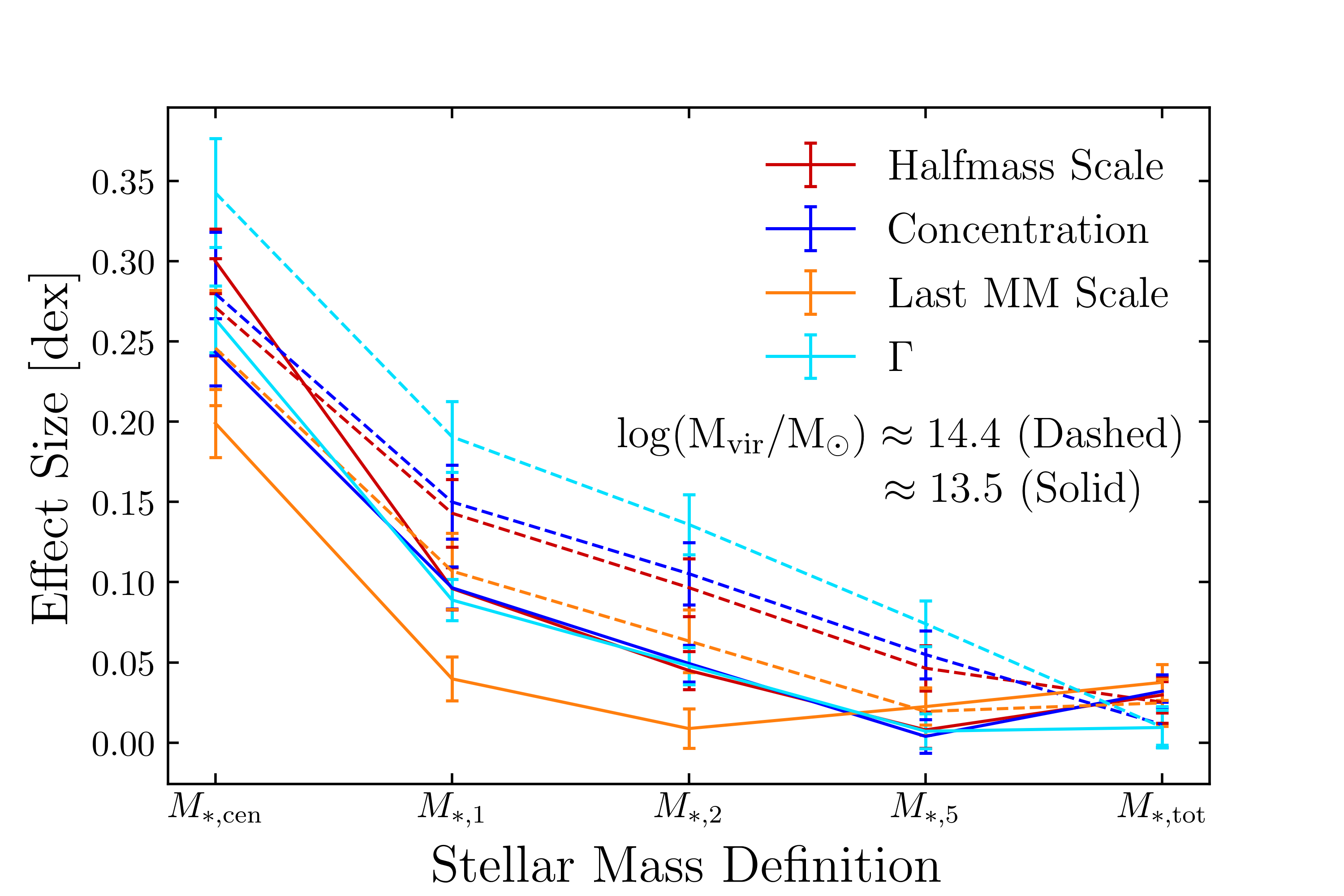}
    \caption{\
        The effect that selected secondary halo properties have on various stellar masses.
        At fixed \Mhalo{}, the secondary property effect size is defined as the difference in stellar mass between halos in the top and bottom 20\% of that secondary parameter. This can be visualized as the length of the vertical gray lines in \autoref{fig:age_split_smhm}.
        Stellar mass definitions that include more of the stellar mass are less affected by secondary halo properties, and \M{tot} is almost uncorrelated with all tested secondary properties.
        However, at high masses more satellites are needed to reduce the effect size than at lower masses.
        \label{fig:effect_size}
    }
    \end{center}
\end{figure}

Recent work has argued that \M{cen} depends both on \Mhalo{} and on the halo assembly history \citep[\eg{}][]{Hearin2013, Rodriguez-Puebla2015, Hearin2015a, Saito2016, Matthee2017}.
If \M{cen} does depend on properties other than \Mhalo{}, and those properties have variance at fixed \Mhalo{}, this could be a source of the scatter in \M{cen}.
In this section we show that, in the UM, \M{cen} depends strongly on secondary properties, while \M{tot} does not. We discuss the implications of this result for both observational measurements and simple mock making techniques such as abundance matching.

The left panel of \autoref{fig:age_split_smhm} shows that, in the UM, \M{cen} depends strongly on halo age. At fixed \Mhalo{}, halos that formed earlier contain larger centrals than those that formed later.
The variance of halo age at fixed mass is therefore a source of some of the intrinsic scatter in the \M{cen} -- \Mhalo{} relation.

The age dependence also means that an \M{cen} cut preferentially selects old halos.
\autoref{table:mcen_shifts} demonstrates this by comparing the median value of various secondary halo properties in two samples.
The first sample is selected by a thin \M{cen} cut, while the second is selected randomly to match the first's \Mhalo{} distribution.
The \M{cen} selected sample has secondary properties biased in a direction that indicates earlier halo formation (less recent accretion, a higher concentration, an earlier halfmass age) than the random sample.

\begin{table}
\begin{tabular}{l l l}
    \toprule
    Characteristic              & \M{cen} selected  & \Mhalo{} matched  \\ \midrule
    Concentration               & 5.60              & $4.82  \pm 0.09$  \\ \midrule
    Halfmass Scale              & 0.450             & $0.491 \pm 0.003$ \\ \midrule
    Last MM Scale               & 0.353             & $0.445 \pm 0.013$ \\ \midrule
    Acc. Rate ($\Gamma$)        & 0.753             & $0.939 \pm 0.056$ \\ \bottomrule
\end{tabular}
\caption{\
    The median value of secondary halo characteristics for two selections of halos. The first is selected with a thin \M{cen} cut ($11.9 < \tlog(\M{cen} / \Msol{}) < 12.1$). The second is selected randomly, but weighted to have the same \Mhalo{} distribution as the first.
    Despite having the same halo mass distributions, the two selections of halos have significantly different assembly histories.
    As shown in \autoref{fig:age_split_smhm}, the \M{cen} selection is consistent with being older.
    \label{table:mcen_shifts}
}
\end{table}

In contrast, the right panel of \autoref{fig:age_split_smhm} shows that the impact of age on the \M{tot} -- \Mhalo{} relation is negligible.
As a result, there is little contribution to the scatter from age, and, as \autoref{table:mtot_shifts} shows, a sample selected by an \M{tot} cut is similar to one selected with a matching \Mhalo{} distribution.

\begin{table}
\begin{tabular}{l l l}
    \toprule
    Characteristic              & \M{tot} selected  & \Mhalo{} matched   \\ \midrule
    Concentration               & 4.68              & $4.73  \pm 0.08$   \\ \midrule
    Halfmass Scale              & 0.495             & $0.493 \pm 0.003$  \\ \midrule
    Last MM Scale               & 0.465             & $0.456 \pm 0.013$  \\ \midrule
    Acc. Rate ($\Gamma$)        & 1.036             & $1.015 \pm 0.063$  \\ \bottomrule
\end{tabular}
\caption{\
    The median value of secondary halo characteristics for two selections of halos. The first is selected with a thin \M{tot} cut ($12.38 < \tlog(\M{tot} / \Msol{}) < 12.62$) chosen to match the number density of the \M{cen} cut used in \autoref{table:mcen_shifts}. The second is selected randomly, but weighted to have the same \Mhalo{} distribution as the first.
    These selections have similar secondary characteristics.
    \label{table:mtot_shifts}
}
\end{table}

We now test how the cen+N mass is affected by secondary properties.
We quantify the effect of a secondary property as the difference in stellar mass at fixed \Mhalo{} for halos in the top and bottom 20\% of that property (\ie{} the gap between the two lines in \autoref{fig:age_split_smhm} highlighted by the vertical gray lines).
\autoref{fig:effect_size} shows how the effect varies for a number of secondary halo properties as the stellar mass definition changes.
We find that increasing the number of satellites included in the stellar mass decreases the dependence on all tested secondary parameters.
As with the scatter, the number of satellites required to achieve results comparable to \M{tot} depends on halo mass.
At $\tlog{}\,\Mhalo{} / \Msol{} \approx 13.5$, there is little gain from using satellites beyond the most massive two as the effect size for \M{2} is comparable to that for \M{tot}.
At $\tlog{}\,\Mhalo{} / \Msol{} \approx 14.4$, the effect size of \M{5} is not quite comparable to that of the total stellar mass, though, as mentioned before, these larger halos will contain more bright satellites, and for \M{5} the dependence is only 0.05 \dex{}.

The fact that \M{tot} does not depend on secondary halo properties, and that this dependence is reduced for the cen+N mass (\eg{} \M{2}, \M{5}), make these promising observables to use to identify and estimate the mass of halos.
Not only do these estimates suffer from less intrinsic scatter, but they also select populations that are less biased with respect to halo assembly.
This is also useful when making mocks as these properties can be accurately assigned using \Mhalo{}, without considering secondary terms.

A benefit of having one observable that is correlated with secondary properties (\M{cen}) and another that is not (\M{tot}) is discussed in Xhakaj \etal{} (in prep.), which shows how observations of \M{cen} and \M{tot} can be combined to form a proxy for halo age.

\section{Physical Origin of \scatterMstarcentitle{} in Groups and Clusters}\label{sec:origin}

We have shown that \M{cen} has both a larger scatter and a greater dependence on halo assembly than \M{tot}.
However, in massive halos, \M{cen} is dominated by its \exsitu{} component and therefore, like \M{tot}, grows primarily through mergers.
This suggests that the growth of \M{cen} can be thought of as a two-stage process, where stellar mass is first brought into the halo,  and then subsequently deposited onto the central.
In this section, we decompose \scatterMstarx{cen} into separate components relating to these two stages.
The first, \scatterMstarx{tot}, is shown to be stochastic.
The second, \scatterCenTot{}, depends on halo age.

We emphasize that the conclusions of this section are only valid in the regime where \M{cen} is dominated by accreted mass (\M{ex}) rather than star formation (\M{in}), which is true in the UniverseMachine for halos with $\FullMhalo > 14$.

\subsection{Decomposition of \scatterMstarx{cen}}

The central stellar mass is comprised of two components,

\begin{equation}
\M{cen} = \M{in} + \M{ex}
\end{equation}

\begin{figure}
    \begin{center}
    \includegraphics[width=0.5\textwidth]{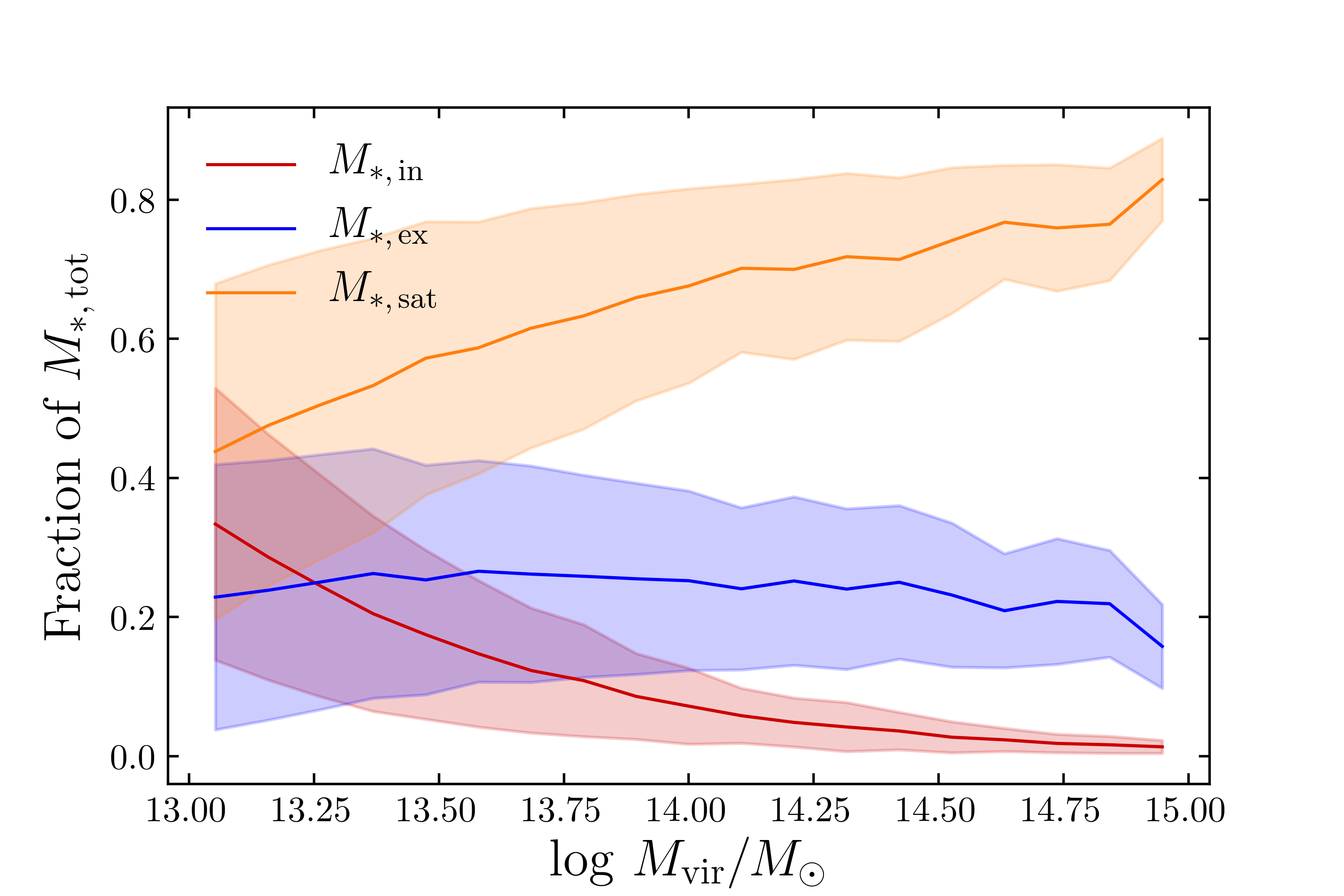}
    \caption{\
        The distribution of \M{tot} as a function of halo mass.
        On average, in halos with $\FullMhalo \approx 14$, one third of the stellar mass is in the central and two thirds in satellites. Of \M{cen}, 80\% comes from mergers (\M{ex}) and 20\% from \insitu{} star formation (\M{in}). At higher masses, the fraction of mass in satellites increases, while the fraction of \M{in} decreases.
        \label{fig:wheres_the_mass}
    }
    \end{center}
\end{figure}

\noindent However, \autoref{fig:wheres_the_mass} shows that for large halos \M{ex} dominates: in halos with $\FullMhalo > 14$, the \exsitu{} fraction is $> 80\%$.
This is consistent with the \exsitu{} fraction of the most massive galaxies in hydrodynamical simulations \citep[\eg{}][]{Lee2013, Pillepich2017}.
For halos in this mass range, we can therefore neglect the \insitu{} component and approximate $\M{cen} \approx \M{ex}$.

The accretion of stellar mass onto the central (\exsitu{} growth) can be thought of as a two-stage process. First, stellar mass is brought into the halo in satellite galaxies. Second, those satellites merge onto the central.
Both of these stages have some scatter, which we assume is lognormal. In the first stage, \M{tot} is a function of \Mhalo{},

\begin{equation}
    \log \M{tot} \sim{} \mathcal{N}(\Mbar{tot}(\Mhalo{}), \scatterMstarx{tot})
\end{equation}

\noindent while in the second stage, \M{cen} is a function of \M{tot}: it is the fraction of \M{tot} that has merged onto the central,

\begin{equation}
    \log \M{cen} \sim{} \mathcal{N}(\Mbar{cen}(\M{tot}), \scatterCenTot{})
\end{equation}

\noindent We can therefore express $\M{cen}$ as a function of $\Mhalo{}$ by combining these two stages,

\begin{align}
    \log \M{cen}& \sim{} \mathcal{N}( \\
         &\Mbar{cen}(\mathcal{N}(\Mbar{tot}(\Mhalo{}), \scatterMstarx{tot}), \\
         &\scatterCenTot)
\end{align}

\noindent We find that, in the UM, the two scatters (\scatterMstarx{tot}, \scatterCenTot{}) are uncorrelated (their correlation coefficient is $< 0.02$). Because of this, the overall scatter in \M{cen} at fixed \Mhalo{} is described by,

\begin{equation}
    \scatterMstarx{cen}^2 = (\frac{d\Mbar{cen}}{d\M{tot}} \scatterMstarx{tot})^2 + \scatterCenTot^2
\end{equation}

We now consider the physical processes that drive these two components.

\subsection{Stochastic component (\scatterMstarx{tot})}

We have already shown in \autoref{sec:biases} that the total stellar content of halos does not correlate with other halo properties (see \autoref{fig:age_split_smhm} and \autoref{fig:effect_size}).
We now show that the scatter is consistent with being stochastic.
We also show that the decrease in \scatterMstarx{tot} with halo mass, as shown in \autoref{fig:intro_plot}, is the result of the statistics of hierarchical assembly.
We present two simple experiments that illustrate these points.

\subsubsection{Stochasticity test}

While we cannot prove that \M{tot} depends solely on halo mass, we can show that its scatter is similar to that predicted by a simple random model, the conditional luminosity function (CLF) \citep{Yang2003}.
The CLF describes the expected galaxy population of a halo given its mass, $\Phi(\Mstar{} | \Mhalo{}) d\Mstar{}$.
The variant we use, \citep[\eg{}][]{Cooray2006, Lan2016}, consists of central ($\Phi_{c}$) and satellite ($\Phi_{s}$) components.
These distributions can be used to construct simulated cluster galaxies by drawing once from $\Phi_{c}$, and $n$ times from the normalized $\Phi_{s}$ where $n$ is a draw from the Poisson distribution with a mean of the expected number of satellites.
We note that the CLF method simplifies the galaxy-halo relation \citep[\eg{}][]{Zentner2014} and so will not generate perfectly realistic realizations of \M{tot}.
However, it will give a good estimate of the scatter in \M{tot} that is expected from random assembly, as the components are drawn independently from the population.

\begin{figure}
    \begin{center}
    \includegraphics[width=0.5\textwidth]{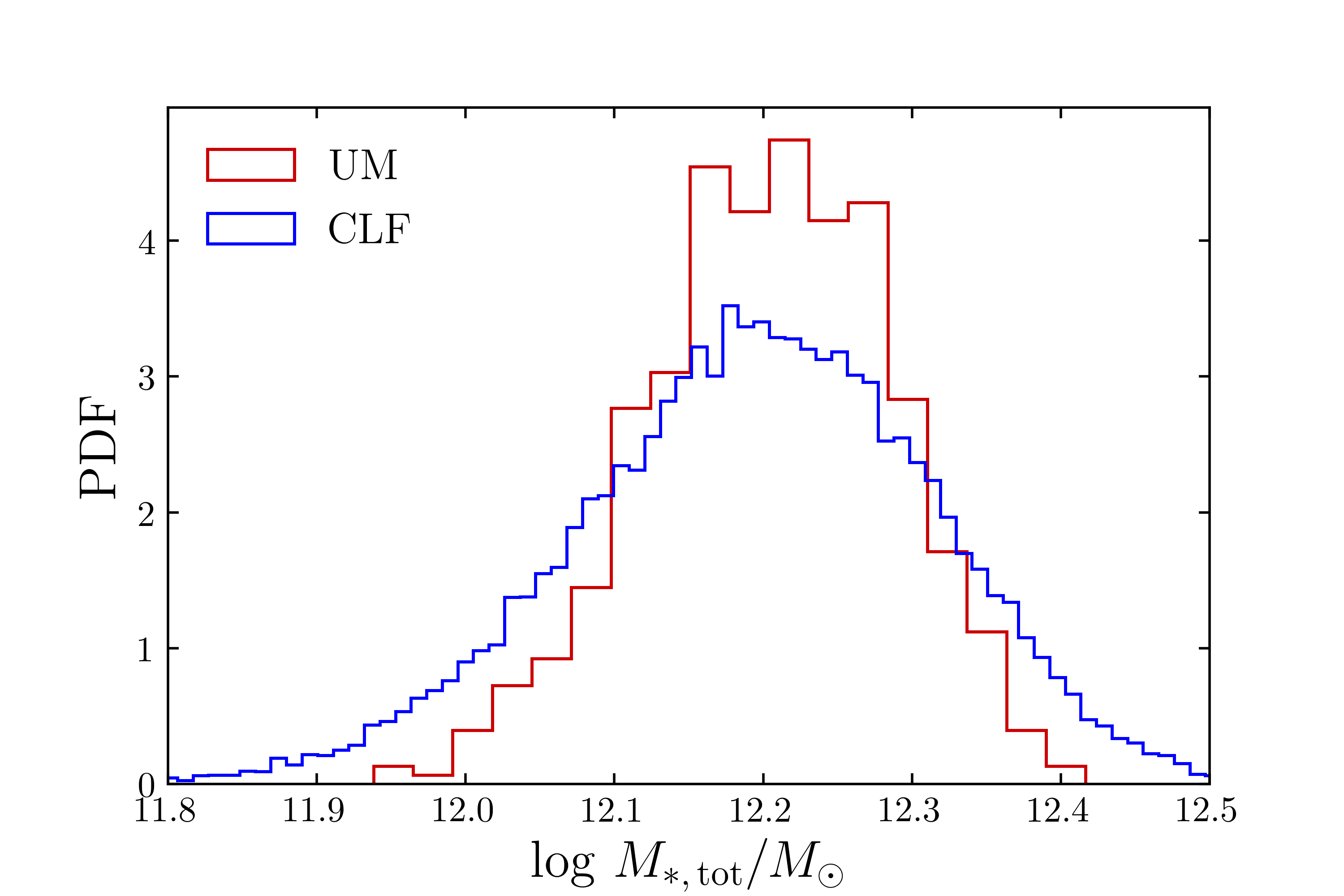}
    \caption{\
        The distribution of \M{tot} for halos with $14 < \FullMhalo{} < 14.1$ in the UM and CLF simulation.
        The distributions are centered within $0.01\ \dex{}$ of each other, and \scatterMstarx{tot} is $0.08$ and $0.12\ \dex{}$ for the UM and CLF respectively.
        The CLF shows the approximate scatter for random assembly and is similar to the scatter in the UM\@.
        \label{fig:CLF_sat_resample}
    }
    \end{center}
\end{figure}

We construct a CLF by setting $\Phi_{c}$, $\Phi_{s}$, and $n$ to be that of the UM in the mass range $14 < \FullMhalo{} < 14.1$.
The comparison between the \M{tot} of UM halos in that mass range and of CLF simulated halos is shown in \autoref{fig:CLF_sat_resample}.
Both distributions are centered at the same \M{tot} to within $0.01\ \dex{}$, and the UM and CLF have \scatterMstarx{tot} of $0.08$ and $0.12\ \dex{}$ respectively.
The slightly larger scatter in the CLF is expected as in the UM the central stellar mass is anticorrelated with that of the satellites.
However, the fact that the \scatterMstarx{tot} in the UM is similar to that of the random CLF suggests that scatter from hierarchical assembly is well modeled by a stochastic process.

\subsubsection{Decreasing scatter expected under stochastic assembly}\label{subsec:decreasing_scatter_explained}

\autoref{fig:intro_plot} showed \scatterMstarx{tot} decreasing significantly with halo mass. We show here that, assuming \M{tot} is determined independently for each halo \ie{} there is no strong environment dependence, this decrease is just a result of the statistical properties of the sum of draws from a lognormal distribution.

Consider the simple case where we assume that clusters are built from a population of progenitors with a single halo mass, $\Mhalo{}_{\rm, prog}$, and so have stellar mass given by $\log \M{tot, prog} \sim \mathcal{N}(\mu, \sigma)$, $\sigma \ll 1$. The cluster that results from the merger of $n$ of these progenitors will have \M{tot} of

\begin{equation}
    \tlog{}(\M{tot}) \approx \tlog{}(n) + \mu \pm \frac{\sigma}{\sqrt{n}}
    \label{eq:scatter_decrease}
\end{equation}

\noindent This is entirely due to the statistics of summing draws from a lognormal distribution, but shows that the lognormal scatter in \M{tot} is expected to decrease with the number of mergers and therefore the mass of the cluster.

\begin{figure}
    \begin{center}
    \includegraphics[width=0.5\textwidth]{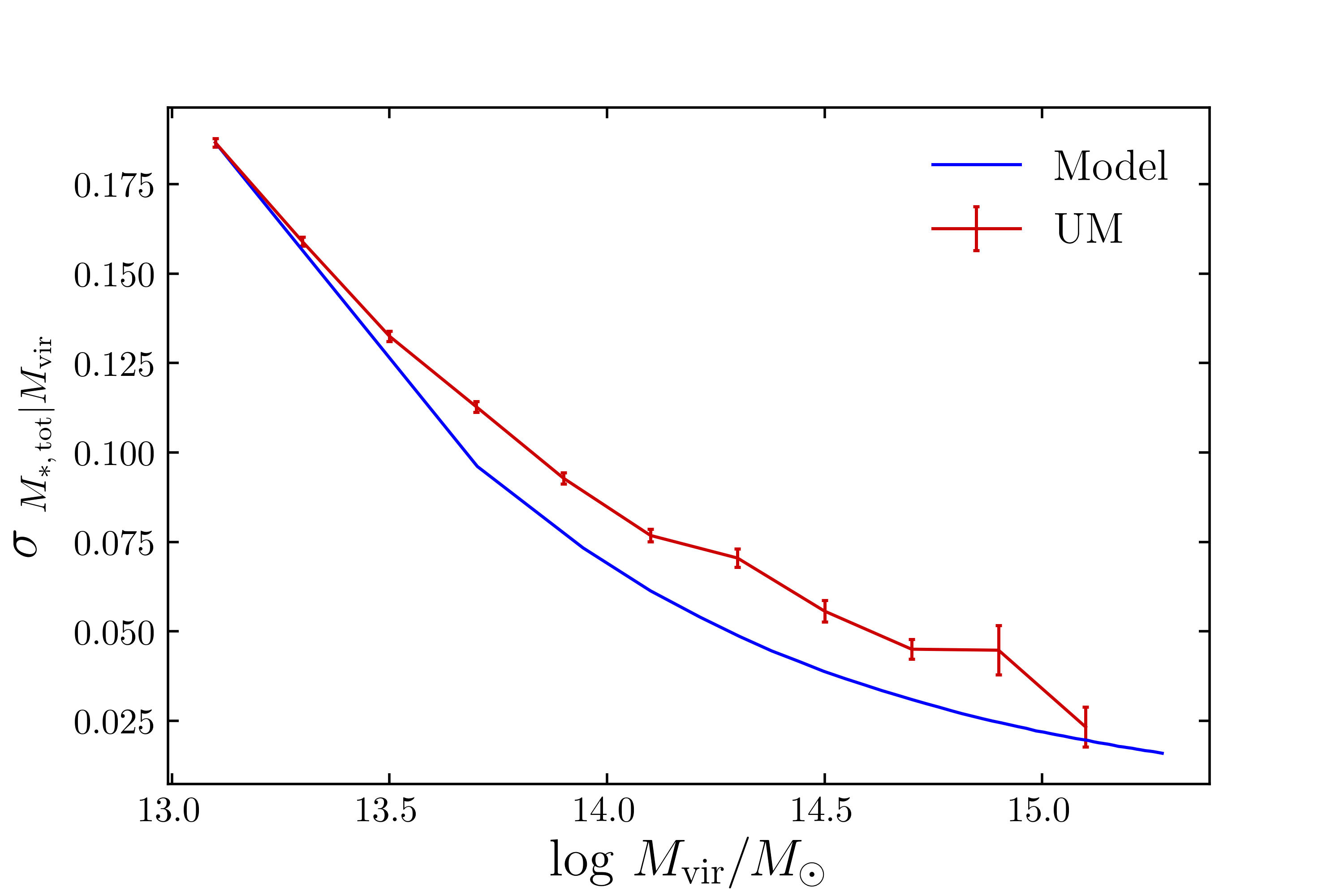}
    \caption{\
        \scatterMstarx{tot} in the UM compared to that of a stochastic toy model.
        The model builds high mass halos from a single mass progenitor population of $\FullMhalo{} = 13.1$ with $\scatterMstarx{tot} \approx 0.18$ (chosen to match the low mass end of the plot).
        The model illustrates that, under the assumption of random assembly, scatter decreases as halo mass increases.
        As the model is an obvious oversimplification (\ie{} single mass progenitors), we do not expect it to be consistent with the UM\@.
        However, the trends are remarkably similar, suggesting that the simple theory of assembly may be appropriate.
        \label{fig:simple_model}
    }
    \end{center}
\end{figure}

In \autoref{fig:simple_model}, we model \scatterMstarx{tot} using \autoref{eq:scatter_decrease} with progenitors of $\FullMhalo = 13.1$ and $\scatterMstarx{tot, prog} = 0.18$ (chosen to fit the low mass end of the UM). The predictions of scatter at the high mass end, even in this highly simplified model, are relatively accurate.
This simple model ignores contributions to the scatter from varying progenitors masses, and variance across halos in the unevolved progenitor mass function. However, the added scatter from this second component will be limited by the universality of the mass function \citep[\eg{}][]{Jiang2016}.

While this section is primarily concerned with \scatterMstarx{tot}, as \M{cen} for massive halos is dominated by its \exsitu{} component, the same statistical argument explains why \scatterMstarx{cen} also decreases with halo mass.
A more detailed analysis of the effect of mergers on \scatterMstarx{cen} (and applicable to \scatterMstarx{tot}) is the Monte Carlo simulation shown in Figure 6 of \citet[][]{Gu2016}. This model assumes no initial scatter in the stellar to halo mass relation, so scatter initially increases with mergers.
However, as in our model, after a large number of mergers, the scatter tends asymptotically to 0.

\subsubsection{Conclusions}

These two experiments suggest a physical cause for \scatterMstarx{tot} in clusters.
We assume that early galaxies have a lognormally distributed \M{tot}, with the scatter primarily due to variations in the central star formation efficiency.
Cluster mass halos then assemble by random mergers of many of these early-galaxy progenitors.
This simple picture {\em generically\/} predicts a \scatterMstarx{tot} that decreases with increasing halo mass.\footnote{As a corollary, we note that the same statistical argument leads to the generic decrease in the scatter \scatterMstarx{cen} with halo mass, since \M{cen} is dominated by \M{ex} in cluster-mass halos.}

\subsection{Age dependent component (\scatterCenTot{})}\label{subsec:origin_scatter_cen_tot}

We have shown in \autoref{sec:biases} that \M{cen} depends both on halo mass and assembly, with halos that formed earlier hosting a larger central galaxy.
We now show that the cause of this age dependence is the \exsitu{} component of \M{cen}, which dominates at high masses.
We then determine how much of \scatterCenTot{} can be explained by halo properties that summarize assembly.

\subsubsection{Source of \scatterCenTot{}}

\begin{figure}
    \begin{center}
    \includegraphics[width=0.5\textwidth]{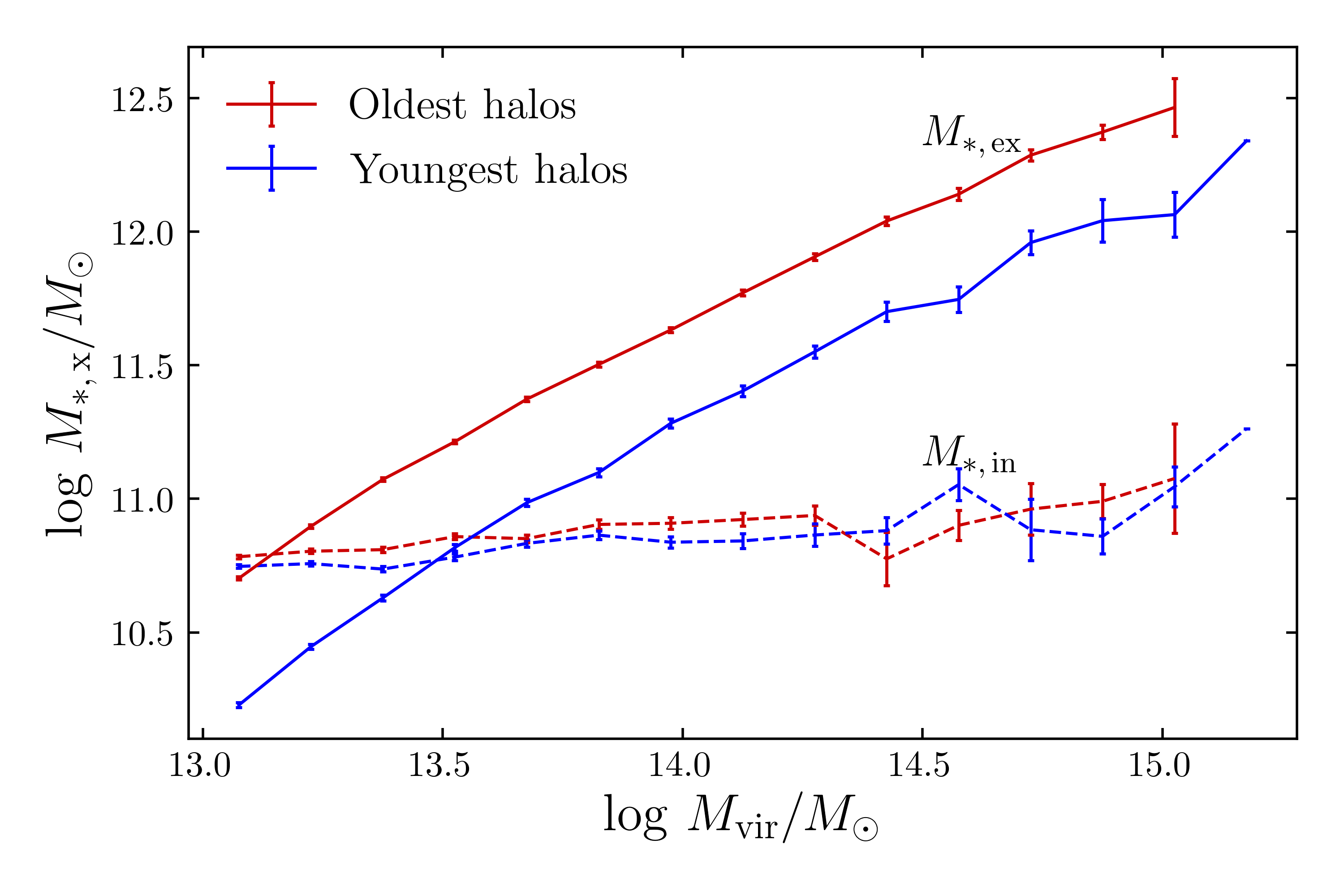}
    \caption{\
        The \M{ex} -- \Mhalo{} and \M{in} -- \Mhalo{} relations for the youngest and oldest 20\% of halos in each mass bin.
        \M{ex} depends strongly on age: the \exsitu{} mass in the oldest quintile of halos is $\mysim{} 0.5$ \dex{} larger than in the youngest.
        In older halos, more stellar mass from mergers has deposited onto the central galaxy and less exists in satellites.
        While the difference between the mean \M{in} in the oldest and youngest halos is statistically significant, as this difference is $< 0.05$ \dex{} it is not practically significant.
        \label{fig:age_split_in_ex}
    }
    \end{center}
\end{figure}

\autoref{fig:age_split_in_ex} shows how the \insitu{} and \exsitu{} components of \M{cen} depend on halo age. The mass of \insitu{} stars is nearly independent of age, but the \exsitu{} component of the oldest quintile of halos is $\mysim{} 0.5$ \dex{} larger than that of the youngest.
Age therefore primarily influences the stellar mass accreted onto the central from mergers, {\em not\/} the stellar mass formed in the central \insitu{}.
In older halos, more stellar mass has been deposited onto the central galaxy, whereas in recently formed halos, this additional mass is still bound up in satellites.

To what extent can \scatterCenTot{} be understood in terms of the dependence of \M{ex} on halo age?
To test this, we define $\deltaCenTot{} = \M{cen} - \Mbar{cen}(\M{tot})$, the difference between the actual \M{cen} and the expected \M{cen} given \M{tot}.
Halos with larger (smaller) central galaxies than expected, given their \M{tot}, will therefore have positive (negative) \deltaCenTot{}.
We then use a Spearman rank correlation to find the halo properties that correlate best with \deltaCenTot{} and therefore the halo properties that most strongly predict deviations from the mean relation.
The Spearman rank correlation coefficient describes how well a monotonic function could fit the data, with $R_s =$ 1 (-1) indicating a perfect, increasing (decreasing) correlation and $R_s = 0$, no correlation. \autoref{table:spearman_cen_halo} shows the halo properties with the most significant Spearman correlations with \deltaCenTot{}. These properties are all correlated with the halo formation time.

\begin{table}
\begin{tabular}{l  l}
    \toprule
    Characteristic              & $R_s$     \\ \midrule
    Halfmass Scale              & -0.56     \\ \midrule
    Concentration               &  0.46     \\ \midrule
    Accretion Rate ($\Gamma$)   & -0.51     \\ \midrule
    Last MM Scale               & -0.41     \\ \midrule
\end{tabular}
\caption{\
    The Spearman rank correlation coefficients for halo properties that correlate best with \deltaCenTot{}, the difference in dex between the true \M{cen} and the expected \M{cen} given \M{tot}.
    For halo properties where a large value implies that most growth happened at late times (\eg{} accretion rate, last major merger scale, halfmass scale), there is a negative correlation:
    halos whose growth happened at late times have smaller than expected, given \M{tot}, central galaxies.
    A large concentration, indicative of early halo formation \citep[\eg{}][]{Wechsler2002}, implies a larger central, given total stellar mass.
    \deltaCenTot{} is therefore sensitive to the halo formation time.
    \label{table:spearman_cen_halo}
}
\end{table}

\begin{figure}
    \begin{center}
    \includegraphics[width=0.5\textwidth]{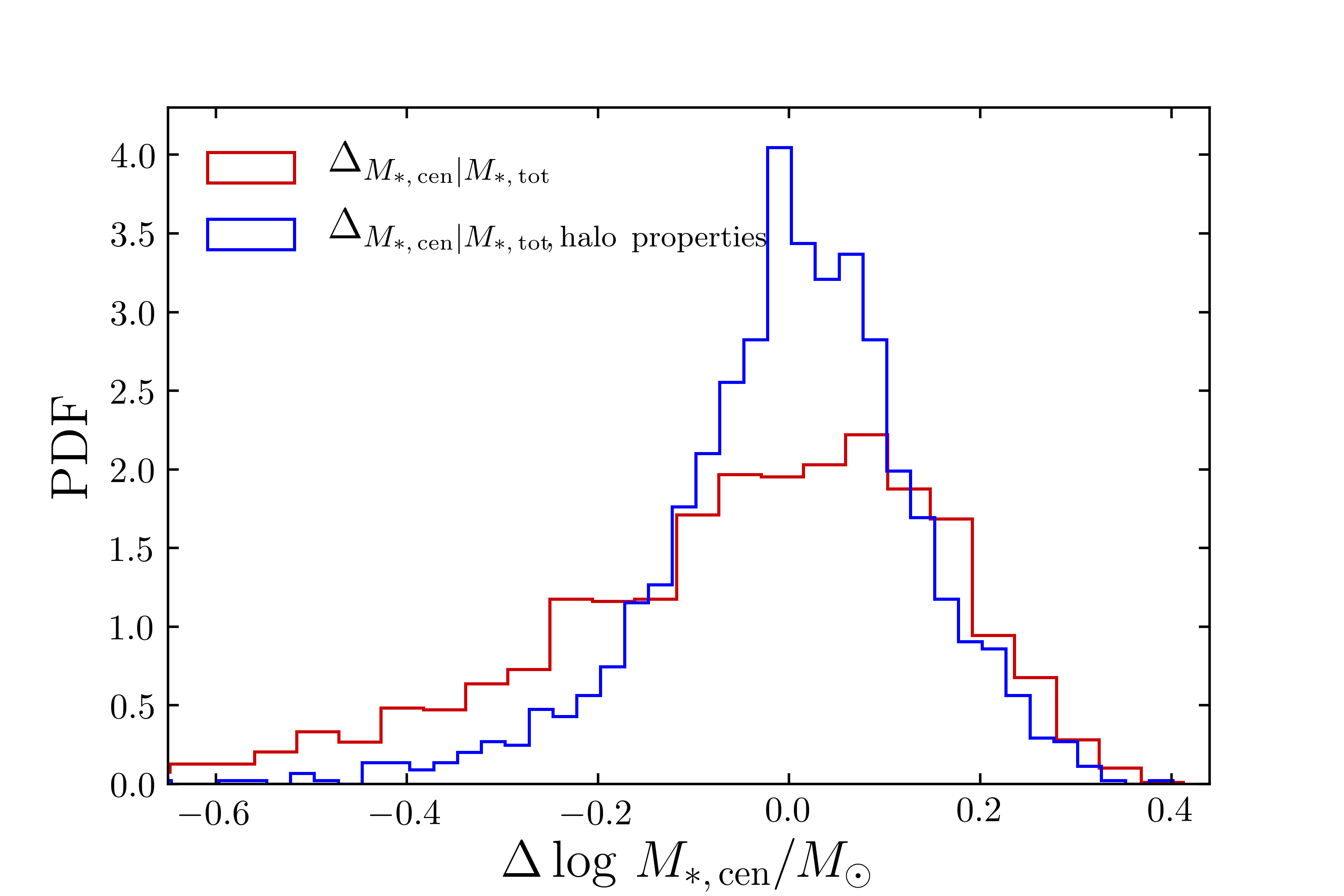}
    \caption{\
        The scatter in \M{cen} is significantly reduced if secondary halo properties (the most important of which are shown in \autoref{table:spearman_cen_halo}) are also controlled for.
        This plot compares the scatter in \M{cen} first at fixed \M{tot} (\deltaCenTot{}), and second at fixed \M{tot} and with a linear model to control for secondary halo properties (\deltaCenTotHP{}).
        The scatter is significantly reduced in the second case, from $0.19$ to $0.13$ \dex{}.
        This is for cluster mass ($\tlog\, \Mhalo{} / \Msol{} > 14$) halos.
        \label{fig:improvement_mcen_mhalo}
    }
    \end{center}
\end{figure}

We build a model for \M{cen} that includes both \M{tot} and the halo properties that have a strong correlation with \deltaCenTot{}.
The model for the halo properties is a linear regression as we find that allowing higher order polynomial fits does not significantly improve the performance.
\autoref{fig:improvement_mcen_mhalo} compares the scatter in \M{cen} using just \M{tot} and using both \M{tot} and other halo properties.
Including the halo properties reduces scatter from $0.19$ to $0.13$ \dex{}: these properties can account for a significant fraction, but not all, of \scatterCenTot{}.

The large residual scatter is not surprising.
First, the halo properties we used in our model are very broad summaries of the mass accretion history. They do not capture details such as the merger orbital parameters and mass ratios, both of which dramatically affect the time it takes for satellites to deposit onto the central \citep{Boylan-Kolchin2007, Boylan-Kolchin2008}.
Second, the choice of star formation rates in the UM has some intrinsic randomness to allow it to mimic scatter caused by physical processes that are not captured in N-body simulations.
An example of this is baryonic effects, which recent work with hydrodynamic simulations has shown can also explain some of the scatter \citep[\eg{}][]{Matthee2017, Kulier2018}.
However, it is clear that age is a major component of \scatterCenTot{}, and therefore of \scatterMstarx{cen}.

\subsubsection{Implications for richness-based halo mass estimates}

The fact that \M{cen} increases, and therefore the stellar mass in satellites decreases, with halo age is a concern for richness-based mass estimators.
It suggests that, at fixed \Mhalo{}, halos that assembled later will have a higher richness.
Here we show that richness, measured by our proxy \ngals{}, is indeed influenced by halo age, albeit with a significantly weaker correlation relative to \M{cen}.

\begin{figure}
    \begin{center}
    \includegraphics[width=0.5\textwidth]{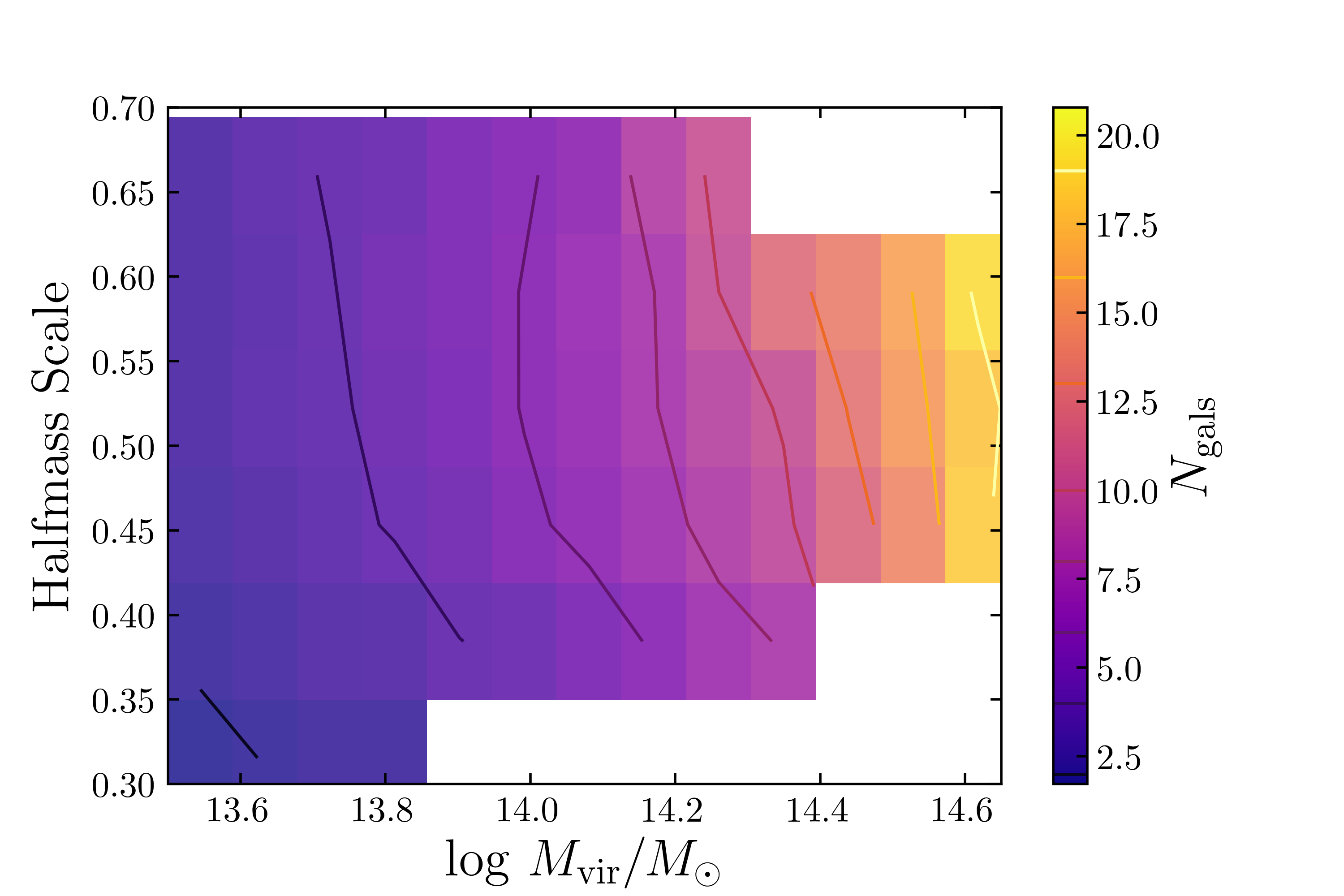}
    \caption{\
        Richness, measured by the proxy \ngals{}, as a function of \Mhalo{} and the halfmass scale ($a$).
        The contours of fixed richness are not vertical and therefore the age of the halo can influence richness-based \Mhalo{} estimates.
        The effect of age is $\sim 0.1$ \dex{}, for example $\ngals{} = 10$ is the mean richness for both $\Mhalo{} = 14.4$, $a = 0.4$ and $\Mhalo{} = 14.25$, $a = 0.65$.
        The uncertainty on the mean richness (the contours) is negligible at all but the largest halo masses.
        \label{fig:gamma_mvir_and_richness}
    }
    \end{center}
\end{figure}

\autoref{fig:gamma_mvir_and_richness} shows the contours of constant \ngals{} as a function of \Mhalo{} and halfmass scale ($a$).
As the contours are not vertical, richness is a function both of halo mass and halo age. At fixed halo mass older halos are less rich than younger ones, an effect that can add a scatter of $\sim 0.1$ \dex{} in \Mhalo{} at fixed $\ngals{}$.
The uncertainties are not shown in the plot, but while the scatter in richness within each bin of \Mhalo{} and $a$ is relatively large ($\sim 1$ and $3$ at the low and high mass ends respectively), the uncertainty on the mean richness is negligible at all but the highest masses.

The dependence of $\ngals{}$ on age is, however, significantly smaller than that of \M{cen}.
The dynamics of mergers offers a natural explanation for this.
It is well known that it takes less time for larger satellite galaxies to merge with the central than smaller ones \citep[\eg{}][]{Lacey1993, Boylan-Kolchin2007, Jiang2008}.
The mass of the central will, after a relatively short period of time, be significantly boosted by these large mergers.
On the other hand, the richness is dominated by galaxies just above the $\ngals{}$ mass cutoff.
These smaller galaxies will take longer to merge onto the central, and so the effect of age on richness is smaller than on \M{cen}.

\section{Summary and Conclusions}\label{sec:conclusions}

Obtaining accurate halo mass estimates is essential for constraining cosmology, but is made difficult by scatter in the observable -- \Mhalo{} relation.
With an eye to upcoming large optical surveys, we have used the UniverseMachine to investigate the causes of the scatter of two halo mass proxies, \M{cen} and \M{tot}.
We have also proposed a new low scatter observable, the cen+N mass (\M{N}), defined as the sum of the stellar mass of the central galaxy and the $N$ most massive satellites.
Our main results are summarized below.

\begin{itemize}
    \item We introduced the cen+N mass, the sum of the central and the $N$ most massive satellites (\M{N}), and showed that its scatter is significantly smaller than that of \M{cen} and approaches, with relatively small $N$, that of \M{tot}.
        We find that the cen+N mass has intrinsic scatter comparable to richness-based estimators with only a few (2 -- 5) of the most massive satellites used.
        However, we showed that it performs better than richness-based estimators under simple tests of projection.
    \item For all definitions of stellar mass (\M{cen}, \M{tot}, \M{N}), the scatter decreases significantly with halo mass. While most previous works have assumed mass independent scatter, we show that this decrease is expected from the statistics of hierarchical assembly.
    \item We find that \M{tot} is a function only of \Mhalo{} with some scatter. We show that this scatter exhibits no correlation with secondary halo properties and is consistent with being stochastic.
    \item On the other hand, \M{cen} depends {\em both} on \Mhalo{} and the halo's age. Moreover, this dependence is almost entirely due to the size of the \exsitu{} component of \M{cen}: In halos with early formation times, a large fraction of stellar mass from mergers has been deposited onto the central; in halos with late formation times, more of this mass is bound up in satellites.
        Insofar as the UniverseMachine model accurately approximates how massive centrals are assembled in the real universe, this implies that any \M{cen}-based selection of massive galaxies will be biased towards samples residing in old halos.
\end{itemize}

We made a number of simplifications and assumptions in this work.
Most importantly, our entire analysis is predicated on the UniverseMachine.
We also assumed that stellar masses are known with no uncertainty, while in practice these need to be inferred from observations of the luminosity. These observations will include some scatter and will likely not include all light.
Apart from a single section, we also assumed that we have perfect information about cluster membership.
However, with these assumptions, we have shown that the cen+N mass is a viable, low-scatter, minimally biased, halo mass proxy that appears to perform well with spectroscopic data.
In future work, we plan to see if similar effects are seen in the IllustrisTNG300 simulations \citep{Nelson2018}, and to apply the cen+N estimator to observations.

\section{Acknowledgments}

This research was supported in part by the National Science Foundation under Grant No. NSF PHY-1748958. This material is based on work supported by the U.D Department of Energy, Office of Science, Office of High Energy Physics under Award Number DE-SC0019301. AL acknowledges support from the David and Lucille Packard foundation, and from the Alfred P. Sloan foundation. The CosmoSim database used in this paper is a service by the Leibniz-Institute for Astrophysics Potsdam (AIP). The MultiDark database was developed in cooperation with the Spanish MultiDark Consolider Project CSD2009-00064.
The authors gratefully acknowledge the Gauss Centre for Supercomputing e.V. (www.gauss-centre.eu) and the Partnership for Advanced Supercomputing in Europe (PRACE, www.prace-ri.eu) for funding the MultiDark simulation project by providing computing time on the GCS Supercomputer SuperMUC at Leibniz Supercomputing Centre (LRZ, www.lrz.de).
The Bolshoi simulations have been performed within the Bolshoi project of the University of California High-Performance AstroComputing Center (UC-HiPACC) and were run at the NASA Ames Research Center.
This research made extensive use the scientific python stack, including:
\href{https://www.numpy.org/}{NumPy} \citep{Oliphant2015},
\href{https://www.numpy.org/}{SciPy} \citep{Jones2001},
\href{https://matplotlib.org/}{Matplotlib} \citep{Hunter2007},
\href{https://pandas.pydata.org/}{Pandas} \citep{McKinney2010},
\href{https://scikit-learn.org/}{Scikit-learn} \citep{Pedregosa2011},
as well as astronomical packages:
\href{https://www.astropy.org}{Astropy} \citep{Robitaille2013, Price-Whelan2018},
\href{http://www.benediktdiemer.com/code/colossus/}{Colossus} \citep{Diemer2018},
\href{https://github.com/astropy/halotools}{Halotools} \citep{Hearin2017}.


\bibliographystyle{mnras}
\bibliography{main}

\begin{thebibliography}{}
\makeatletter
\relax
\def\mn@urlcharsother{\let\do\@makeother \do\$\do\&\do\#\do\^\do\_\do\%\do\~}
\def\mn@doi{\begingroup\mn@urlcharsother \@ifnextchar [ {\mn@doi@}
  {\mn@doi@[]}}
\def\mn@doi@[#1]#2{\def\@tempa{#1}\ifx\@tempa\@empty \href
  {http://dx.doi.org/#2} {doi:#2}\else \href {http://dx.doi.org/#2} {#1}\fi
  \endgroup}
\def\mn@eprint#1#2{\mn@eprint@#1:#2::\@nil}
\def\mn@eprint@arXiv#1{\href {http://arxiv.org/abs/#1} {{\tt arXiv:#1}}}
\def\mn@eprint@dblp#1{\href {http://dblp.uni-trier.de/rec/bibtex/#1.xml}
  {dblp:#1}}
\def\mn@eprint@#1:#2:#3:#4\@nil{\def\@tempa {#1}\def\@tempb {#2}\def\@tempc
  {#3}\ifx \@tempc \@empty \let \@tempc \@tempb \let \@tempb \@tempa \fi \ifx
  \@tempb \@empty \def\@tempb {arXiv}\fi \@ifundefined
  {mn@eprint@\@tempb}{\@tempb:\@tempc}{\expandafter \expandafter \csname
  mn@eprint@\@tempb\endcsname \expandafter{\@tempc}}}

\bibitem[\protect\citeauthoryear{Abazajian et~al.,}{Abazajian
  et~al.}{2009}]{Abazajian2009}
Abazajian K.~N.,  et~al., 2009, \mn@doi [\apjs] {10.1088/0067-0049/182/2/543},
  182, 543–558

\bibitem[\protect\citeauthoryear{Abbott et~al.,}{Abbott
  et~al.}{2018}]{Abbott2018}
Abbott T. M.~C.,  et~al., 2018, \mn@doi [\apjs] {10.3847/1538-4365/aae9f0},
  239, 18

\bibitem[\protect\citeauthoryear{Aihara et~al.,}{Aihara
  et~al.}{2018}]{Aihara2018}
Aihara H.,  et~al., 2018, \mn@doi [\pasj] {10.1093/pasj/psx066}, 70, S4

\bibitem[\protect\citeauthoryear{Andreon}{Andreon}{2012}]{Andreon2012}
Andreon S.,  2012, \mn@doi [\aap] {10.1051/0004-6361/201220284}, 548, A83

\bibitem[\protect\citeauthoryear{Behroozi, Conroy  \& Wechsler}{Behroozi
  et~al.}{2010}]{Behroozi2010}
Behroozi P.~S.,  Conroy C.,   Wechsler R.~H.,  2010, \mn@doi [\apj]
  {10.1088/0004-637X/717/1/379}, 717, 379–403

\bibitem[\protect\citeauthoryear{Behroozi, Wechsler  \& Wu}{Behroozi
  et~al.}{2013a}]{Behroozi2013b}
Behroozi P.~S.,  Wechsler R.~H.,   Wu H.-Y.,  2013a, \mn@doi [\apj]
  {10.1088/0004-637X/762/2/109}, 762, 109

\bibitem[\protect\citeauthoryear{Behroozi, Wechsler, Wu, Busha, Klypin  \&
  Primack}{Behroozi et~al.}{2013b}]{Behroozi2013c}
Behroozi P.~S.,  Wechsler R.~H.,  Wu H.-Y.,  Busha M.~T.,  Klypin A.~A.,
  Primack J.~R.,  2013b, \mn@doi [\apj] {10.1088/0004-637X/763/1/18}, 763, 18

\bibitem[\protect\citeauthoryear{Behroozi, Wechsler, Hearin  \&
  Conroy}{Behroozi et~al.}{2018}]{Behroozi2018}
Behroozi P.,  Wechsler R.,  Hearin A.,   Conroy C.,  2018, arXiv e-prints

\bibitem[\protect\citeauthoryear{Bernardi, Meert, Sheth, Vikram,
  Huertas-Company, Mei  \& Shankar}{Bernardi et~al.}{2013}]{Bernardi2013}
Bernardi M.,  Meert A.,  Sheth R.~K.,  Vikram V.,  Huertas-Company M.,  Mei S.,
    Shankar F.,  2013, \mn@doi [\mnras] {10.1093/mnras/stt1607}, 436, 697–704

\bibitem[\protect\citeauthoryear{Bleem et~al.,}{Bleem et~al.}{2015}]{Bleem2015}
Bleem L.~E.,  et~al., 2015, \mn@doi [\apjs] {10.1088/0067-0049/216/2/27}, 216,
  27

\bibitem[\protect\citeauthoryear{Blumenthal, Faber, Primack  \&
  Rees}{Blumenthal et~al.}{1984}]{Blumenthal1984}
Blumenthal G.~R.,  Faber S.~M.,  Primack J.~R.,   Rees M.~J.,  1984, \mn@doi
  [\nat] {10.1038/311517a0}, 311, 517–525

\bibitem[\protect\citeauthoryear{Boylan-Kolchin \& Ma}{Boylan-Kolchin \&
  Ma}{2007}]{Boylan-Kolchin2007}
Boylan-Kolchin M.,  Ma C.-P.,  2007, \mn@doi [\mnras]
  {10.1111/j.1365-2966.2006.11276.x}, 374, 1227–1241

\bibitem[\protect\citeauthoryear{Boylan-Kolchin, Ma  \&
  Quataert}{Boylan-Kolchin et~al.}{2008}]{Boylan-Kolchin2008}
Boylan-Kolchin M.,  Ma C.-P.,   Quataert E.,  2008, \mn@doi [\mnras]
  {10.1111/j.1365-2966.2007.12530.x}, 383, 93–101

\bibitem[\protect\citeauthoryear{Bryan \& Norman}{Bryan \&
  Norman}{1998}]{Bryan1998}
Bryan G.~L.,  Norman M.~L.,  1998, \mn@doi [\apj] {10.1086/305262}, 495,
  80–99

\bibitem[\protect\citeauthoryear{Busch \& White}{Busch \&
  White}{2017}]{Busch2017}
Busch P.,  White S. D.~M.,  2017, \mn@doi [\mnras] {10.1093/mnras/stx1584},
  470, 4767–4781

\bibitem[\protect\citeauthoryear{Cooray}{Cooray}{2006}]{Cooray2006}
Cooray A.,  2006, \mn@doi [\mnras] {10.1111/j.1365-2966.2005.09747.x}, 365,
  842–866

\bibitem[\protect\citeauthoryear{Croton, Gao  \& White}{Croton
  et~al.}{2007}]{Croton2007}
Croton D.~J.,  Gao L.,   White S. D.~M.,  2007, \mn@doi [\mnras]
  {10.1111/j.1365-2966.2006.11230.x}, 374, 1303–1309

\bibitem[\protect\citeauthoryear{DESI-Collaboration et~al.,}{DESI-Collaboration
  et~al.}{2016}]{DESI2016}
DESI-Collaboration et~al., 2016, arXiv e-prints

\bibitem[\protect\citeauthoryear{Diemer}{Diemer}{2017}]{Diemer2017a}
Diemer B.,  2017, \mn@doi [\apjs] {10.3847/1538-4365/aa799c}, 231, 5

\bibitem[\protect\citeauthoryear{Diemer}{Diemer}{2018}]{Diemer2018}
Diemer B.,  2018, \mn@doi [The Astrophysical Journal Supplement Series]
  {10.3847/1538-4365/aaee8c}, 239, 35

\bibitem[\protect\citeauthoryear{Geha, Blanton, Yan  \& Tinker}{Geha
  et~al.}{2012}]{Geha2012}
Geha M.,  Blanton M.~R.,  Yan R.,   Tinker J.~L.,  2012, \mn@doi [\apj]
  {10.1088/0004-637X/757/1/85}, 757, 85

\bibitem[\protect\citeauthoryear{Golden-Marx \& Miller}{Golden-Marx \&
  Miller}{2018}]{Golden-Marx2018}
Golden-Marx J.~B.,  Miller C.~J.,  2018, \mn@doi [\apj]
  {10.3847/1538-4357/aac2bd}, 860, 2

\bibitem[\protect\citeauthoryear{Gu, Conroy  \& Behroozi}{Gu
  et~al.}{2016}]{Gu2016}
Gu M.,  Conroy C.,   Behroozi P.,  2016, \mn@doi [\apj]
  {10.3847/0004-637X/833/1/2}, 833, 2

\bibitem[\protect\citeauthoryear{Guo, White, Li  \& Boylan-Kolchin}{Guo
  et~al.}{2010}]{Guo2010}
Guo Q.,  White S.,  Li C.,   Boylan-Kolchin M.,  2010, \mn@doi [\mnras]
  {10.1111/j.1365-2966.2010.16341.x}, 404, 1111–1120

\bibitem[\protect\citeauthoryear{Hearin \& Watson}{Hearin \&
  Watson}{2013}]{Hearin2013}
Hearin A.~P.,  Watson D.~F.,  2013, \mn@doi [\mnras] {10.1093/mnras/stt1374},
  435, 1313–1324

\bibitem[\protect\citeauthoryear{Hearin, Watson  \& van~den Bosch}{Hearin
  et~al.}{2015}]{Hearin2015a}
Hearin A.~P.,  Watson D.~F.,   van~den Bosch F.~C.,  2015, \mn@doi [\mnras]
  {10.1093/mnras/stv1358}, 452, 1958–1969

\bibitem[\protect\citeauthoryear{Hearin, Zentner, van~den Bosch, Campbell  \&
  Tollerud}{Hearin et~al.}{2016}]{Hearin2016}
Hearin A.~P.,  Zentner A.~R.,  van~den Bosch F.~C.,  Campbell D.,   Tollerud
  E.,  2016, \mn@doi [\mnras] {10.1093/mnras/stw840}, 460, 2552–2570

\bibitem[\protect\citeauthoryear{Hearin et~al.,}{Hearin
  et~al.}{2017}]{Hearin2017}
Hearin A.~P.,  et~al., 2017, The Astronomical Journal, 154, 190

\bibitem[\protect\citeauthoryear{Hoshino et~al.,}{Hoshino
  et~al.}{2015}]{Hoshino2015}
Hoshino H.,  et~al., 2015, \mn@doi [\mnras] {10.1093/mnras/stv1271}, 452,
  998–1013

\bibitem[\protect\citeauthoryear{Huang et~al.,}{Huang et~al.}{2018}]{Huang2018}
Huang S.,  et~al., 2018, arXiv e-prints

\bibitem[\protect\citeauthoryear{Hunter}{Hunter}{2007}]{Hunter2007}
Hunter J.~D.,  2007, \mn@doi [Computing in Science and Engineering]
  {10.1109/MCSE.2007.55}, 9, 90–95

\bibitem[\protect\citeauthoryear{Ivezic et~al.,}{Ivezic
  et~al.}{2019}]{Ivezic2019}
Ivezic Z.,  et~al., 2019, \mn@doi [The Astrophysical Journal]
  {10.3847/1538-4357/ab042c}, 873, 111

\bibitem[\protect\citeauthoryear{Jiang \& van~den Bosch}{Jiang \& van~den
  Bosch}{2016}]{Jiang2016}
Jiang F.,  van~den Bosch F.~C.,  2016, \mn@doi [\mnras] {10.1093/mnras/stw439},
  458, 2848–2869

\bibitem[\protect\citeauthoryear{Jiang, Jing, Faltenbacher, Lin  \& Li}{Jiang
  et~al.}{2008}]{Jiang2008}
Jiang C.~Y.,  Jing Y.~P.,  Faltenbacher A.,  Lin W.~P.,   Li C.,  2008, \mn@doi
  [The Astrophysical Journal] {10.1086/526412}, 675, 1095–1105

\bibitem[\protect\citeauthoryear{Jones, Oliphant, Peterson  et~al.}{Jones
  et~al.}{2001}]{Jones2001}
Jones E.,  Oliphant T.,  Peterson P.,   et~al., 2001, {SciPy}: Open source
  scientific tools for {Python}, \url {http://www.scipy.org/}

\bibitem[\protect\citeauthoryear{Kaiser}{Kaiser}{1987}]{Kaiser1987}
Kaiser N.,  1987, \mn@doi [\mnras] {10.1093/mnras/227.1.1}, 227, 1–21

\bibitem[\protect\citeauthoryear{Klypin, Yepes, Gottlöber, Prada  \&
  Heß}{Klypin et~al.}{2016}]{Klypin2016}
Klypin A.,  Yepes G.,  Gottlöber S.,  Prada F.,   Heß S.,  2016, \mn@doi
  [\mnras] {10.1093/mnras/stw248}, 457, 4340–4359

\bibitem[\protect\citeauthoryear{Kravtsov, Vikhlinin  \& Nagai}{Kravtsov
  et~al.}{2006}]{Kravtsov2006}
Kravtsov A.~V.,  Vikhlinin A.,   Nagai D.,  2006, \mn@doi [\apj]
  {10.1086/506319}, 650, 128–136

\bibitem[\protect\citeauthoryear{Kravtsov, Vikhlinin  \&
  Meshcheryakov}{Kravtsov et~al.}{2018}]{Kravtsov2018}
Kravtsov A.~V.,  Vikhlinin A.~A.,   Meshcheryakov A.~V.,  2018, \mn@doi
  [Astronomy Letters] {10.1134/S1063773717120015}, 44, 8–34

\bibitem[\protect\citeauthoryear{Kulier, Padilla, Schaye, Crain, Schaller,
  Bower, Theuns  \& Paillas}{Kulier et~al.}{2018}]{Kulier2018}
Kulier A.,  Padilla N.,  Schaye J.,  Crain R.~A.,  Schaller M.,  Bower R.~G.,
  Theuns T.,   Paillas E.,  2018, preprint (\mn@eprint {arXiv} {1805.05349})

\bibitem[\protect\citeauthoryear{Lacey \& Cole}{Lacey \&
  Cole}{1993}]{Lacey1993}
Lacey C.,  Cole S.,  1993, \mn@doi [Monthly Notices of the Royal Astronomical
  Society] {10.1093/mnras/262.3.627}, 262, 627–649

\bibitem[\protect\citeauthoryear{Lan, Ménard  \& Mo}{Lan
  et~al.}{2016}]{Lan2016}
Lan T.-W.,  Ménard B.,   Mo H.,  2016, \mn@doi [\mnras]
  {10.1093/mnras/stw898}, 459, 3998–4019

\bibitem[\protect\citeauthoryear{{Laureijs} et~al.,}{{Laureijs}
  et~al.}{2011}]{Laureijs2011}
{Laureijs} R.,  et~al., 2011, arXiv e-prints, \href
  {https://ui.adsabs.harvard.edu/abs/2011arXiv1110.3193L} {p. arXiv:1110.3193}

\bibitem[\protect\citeauthoryear{Leauthaud, Tinker, Behroozi, Busha  \&
  Wechsler}{Leauthaud et~al.}{2011}]{Leauthaud2011}
Leauthaud A.,  Tinker J.,  Behroozi P.~S.,  Busha M.~T.,   Wechsler R.~H.,
  2011, \mn@doi [\apj] {10.1088/0004-637X/738/1/45}, 738, 45

\bibitem[\protect\citeauthoryear{Leauthaud et~al.,}{Leauthaud
  et~al.}{2012}]{Leauthaud2012}
Leauthaud A.,  et~al., 2012, \mn@doi [\apj] {10.1088/0004-637X/744/2/159}, 744,
  159

\bibitem[\protect\citeauthoryear{Lee \& Yi}{Lee \& Yi}{2013}]{Lee2013}
Lee J.,  Yi S.~K.,  2013, \mn@doi [\apj] {10.1088/0004-637X/766/1/38}, 766, 38

\bibitem[\protect\citeauthoryear{Lin, Stanford, Eisenhardt, Vikhlinin, Maughan
  \& Kravtsov}{Lin et~al.}{2012}]{Lin2012}
Lin Y.-T.,  Stanford S.~A.,  Eisenhardt P. R.~M.,  Vikhlinin A.,  Maughan
  B.~J.,   Kravtsov A.,  2012, \mn@doi [\apjl] {10.1088/2041-8205/745/1/L3},
  745, L3

\bibitem[\protect\citeauthoryear{Mahdavi, Hoekstra, Babul, Bildfell, Jeltema
  \& Henry}{Mahdavi et~al.}{2013}]{Mahdavi2013}
Mahdavi A.,  Hoekstra H.,  Babul A.,  Bildfell C.,  Jeltema T.,   Henry J.~P.,
  2013, \mn@doi [\apj] {10.1088/0004-637X/767/2/116}, 767, 116

\bibitem[\protect\citeauthoryear{Mantz et~al.,}{Mantz et~al.}{2016}]{Mantz2016}
Mantz A.~B.,  et~al., 2016, \mn@doi [\mnras] {10.1093/mnras/stw2250}, 463,
  3582–3603

\bibitem[\protect\citeauthoryear{Marriage et~al.,}{Marriage
  et~al.}{2011}]{Marriage2011}
Marriage T.~A.,  et~al., 2011, \mn@doi [\apj] {10.1088/0004-637X/737/2/61},
  737, 61

\bibitem[\protect\citeauthoryear{Matthee, Schaye, Crain, Schaller, Bower  \&
  Theuns}{Matthee et~al.}{2017}]{Matthee2017}
Matthee J.,  Schaye J.,  Crain R.~A.,  Schaller M.,  Bower R.,   Theuns T.,
  2017, \mn@doi [\mnras] {10.1093/mnras/stw2884}, 465, 2381–2396

\bibitem[\protect\citeauthoryear{McClintock et~al.,}{McClintock
  et~al.}{2019}]{McClintock2019}
McClintock T.,  et~al., 2019, \mn@doi [\mnras] {10.1093/mnras/sty2711}, 482,
  1352–1378

\bibitem[\protect\citeauthoryear{McKinney et~al.}{McKinney
  et~al.}{2010}]{McKinney2010}
McKinney W.,  et~al., 2010, in Proceedings of the 9th Python in Science
  Conference. p. 51–56

\bibitem[\protect\citeauthoryear{More, van~den Bosch, Cacciato, Mo, Yang  \&
  Li}{More et~al.}{2009}]{More2009}
More S.,  van~den Bosch F.~C.,  Cacciato M.,  Mo H.~J.,  Yang X.,   Li R.,
  2009, \mn@doi [\mnras] {10.1111/j.1365-2966.2008.14095.x}, 392, 801–816

\bibitem[\protect\citeauthoryear{Moustakas et~al.,}{Moustakas
  et~al.}{2013}]{Moustakas2013}
Moustakas J.,  et~al., 2013, \mn@doi [\apj] {10.1088/0004-637X/767/1/50}, 767,
  50

\bibitem[\protect\citeauthoryear{Mulroy et~al.,}{Mulroy
  et~al.}{2014}]{Mulroy2014}
Mulroy S.~L.,  et~al., 2014, \mn@doi [\mnras] {10.1093/mnras/stu1387}, 443,
  3309–3317

\bibitem[\protect\citeauthoryear{Murata, Nishimichi, Takada, Miyatake,
  Shirasaki, More, Takahashi  \& Osato}{Murata et~al.}{2018}]{Murata2018}
Murata R.,  Nishimichi T.,  Takada M.,  Miyatake H.,  Shirasaki M.,  More S.,
  Takahashi R.,   Osato K.,  2018, The Astrophysical Journal, 854, 120

\bibitem[\protect\citeauthoryear{Muzzin et~al.,}{Muzzin
  et~al.}{2013}]{Muzzin2013}
Muzzin A.,  et~al., 2013, \mn@doi [\apj] {10.1088/0004-637X/777/1/18}, 777, 18

\bibitem[\protect\citeauthoryear{{Nelson} et~al.,}{{Nelson}
  et~al.}{2018}]{Nelson2018}
{Nelson} D.,  et~al., 2018, arXiv e-prints, \href
  {https://ui.adsabs.harvard.edu/abs/2018arXiv181205609N} {p. arXiv:1812.05609}

\bibitem[\protect\citeauthoryear{Oliphant}{Oliphant}{2015}]{Oliphant2015}
Oliphant T.~E.,  2015, Guide to NumPy, 2nd edn.
CreateSpace Independent Publishing Platform, USA

\bibitem[\protect\citeauthoryear{Pearson, Ponman, Norberg, Robotham  \&
  Farr}{Pearson et~al.}{2015}]{Pearson2015}
Pearson R.~J.,  Ponman T.~J.,  Norberg P.,  Robotham A. S.~G.,   Farr W.~M.,
  2015, \mn@doi [\mnras] {10.1093/mnras/stv463}, 449, 3082–3106

\bibitem[\protect\citeauthoryear{Pedregosa et~al.,}{Pedregosa
  et~al.}{2011}]{Pedregosa2011}
Pedregosa F.,  et~al., 2011, Journal of machine learning research, 12,
  2825–2830

\bibitem[\protect\citeauthoryear{Pillepich et~al.,}{Pillepich
  et~al.}{2017}]{Pillepich2017}
Pillepich A.,  et~al., 2017, \mn@doi [Monthly Notices of the Royal Astronomical
  Society] {10.1093/mnras/stx3112}, 475, 648–675

\bibitem[\protect\citeauthoryear{Planck-Collaboration
  et~al.,}{Planck-Collaboration et~al.}{2016a}]{PlanckCollaboration2016}
Planck-Collaboration et~al., 2016a, \mn@doi [\aap]
  {10.1051/0004-6361/201525830}, 594, A13

\bibitem[\protect\citeauthoryear{Planck-Collaboration
  et~al.,}{Planck-Collaboration et~al.}{2016b}]{PlanckCollaboration2016a}
Planck-Collaboration et~al., 2016b, \mn@doi [\aap]
  {10.1051/0004-6361/201525833}, 594, A24

\bibitem[\protect\citeauthoryear{Price-Whelan et~al.,}{Price-Whelan
  et~al.}{2018}]{Price-Whelan2018}
Price-Whelan A.~M.,  et~al., 2018, \mn@doi [The Astronomical Journal]
  {10.3847/1538-3881/aabc4f}, 156, 123

\bibitem[\protect\citeauthoryear{Reddick, Wechsler, Tinker  \&
  Behroozi}{Reddick et~al.}{2013}]{Reddick2013}
Reddick R.~M.,  Wechsler R.~H.,  Tinker J.~L.,   Behroozi P.~S.,  2013, \mn@doi
  [\apj] {10.1088/0004-637X/771/1/30}, 771, 30

\bibitem[\protect\citeauthoryear{Robitaille et~al.,}{Robitaille
  et~al.}{2013}]{Robitaille2013}
Robitaille T.~P.,  et~al., 2013, \mn@doi [Astronomy \& Astrophysics]
  {10.1051/0004-6361/201322068}, 558, A33

\bibitem[\protect\citeauthoryear{Rodriguez-Gomez et~al.,}{Rodriguez-Gomez
  et~al.}{2016}]{Rodriguez-Gomez2016}
Rodriguez-Gomez V.,  et~al., 2016, \mn@doi [\mnras] {10.1093/mnras/stw456},
  458, 2371–2390

\bibitem[\protect\citeauthoryear{Rodríguez-Puebla, Avila-Reese, Yang, Foucaud,
  Drory  \& Jing}{Rodríguez-Puebla et~al.}{2015}]{Rodriguez-Puebla2015}
Rodríguez-Puebla A.,  Avila-Reese V.,  Yang X.,  Foucaud S.,  Drory N.,   Jing
  Y.~P.,  2015, \mn@doi [\apj] {10.1088/0004-637X/799/2/130}, 799, 130

\bibitem[\protect\citeauthoryear{Rodríguez-Puebla, Behroozi, Primack, Klypin,
  Lee  \& Hellinger}{Rodríguez-Puebla et~al.}{2016}]{Rodriguez-Puebla2016}
Rodríguez-Puebla A.,  Behroozi P.,  Primack J.,  Klypin A.,  Lee C.,
  Hellinger D.,  2016, \mn@doi [\mnras] {10.1093/mnras/stw1705}, 462, 893–916

\bibitem[\protect\citeauthoryear{Rozo \& Rykoff}{Rozo \&
  Rykoff}{2014}]{Rozo2014}
Rozo E.,  Rykoff E.~S.,  2014, \mn@doi [\apj] {10.1088/0004-637X/783/2/80},
  783, 80

\bibitem[\protect\citeauthoryear{Rozo et~al.,}{Rozo et~al.}{2009}]{Rozo2009b}
Rozo E.,  et~al., 2009, The Astrophysical Journal, 708, 645

\bibitem[\protect\citeauthoryear{Rozo, Rykoff, Bartlett  \& Melin}{Rozo
  et~al.}{2015}]{Rozo2015}
Rozo E.,  Rykoff E.~S.,  Bartlett J.~G.,   Melin J.-B.,  2015, \mn@doi [\mnras]
  {10.1093/mnras/stv605}, 450, 592–605

\bibitem[\protect\citeauthoryear{Rykoff et~al.,}{Rykoff
  et~al.}{2012}]{Rykoff2012}
Rykoff E.~S.,  et~al., 2012, \mn@doi [\apj] {10.1088/0004-637X/746/2/178}, 746,
  178

\bibitem[\protect\citeauthoryear{Rykoff et~al.,}{Rykoff
  et~al.}{2014}]{Rykoff2014}
Rykoff E.~S.,  et~al., 2014, \mn@doi [\apj] {10.1088/0004-637X/785/2/104}, 785,
  104

\bibitem[\protect\citeauthoryear{Rykoff et~al.,}{Rykoff
  et~al.}{2016}]{Rykoff2016}
Rykoff E.~S.,  et~al., 2016, \mn@doi [\apjs] {10.3847/0067-0049/224/1/1}, 224,
  1

\bibitem[\protect\citeauthoryear{Saito et~al.,}{Saito et~al.}{2016}]{Saito2016}
Saito S.,  et~al., 2016, \mn@doi [\mnras] {10.1093/mnras/stw1080}, 460,
  1457–1475

\bibitem[\protect\citeauthoryear{Springel}{Springel}{2005}]{Springel2005a}
Springel V.,  2005, \mn@doi [\mnras] {10.1111/j.1365-2966.2005.09655.x}, 364,
  1105–1134

\bibitem[\protect\citeauthoryear{Sunyaev \& Zeldovich}{Sunyaev \&
  Zeldovich}{1970}]{Sunyaev1970}
Sunyaev R.~A.,  Zeldovich Y.~B.,  1970, \mn@doi [\apss] {10.1007/BF00653471},
  7, 3–19

\bibitem[\protect\citeauthoryear{Tinker}{Tinker}{2017}]{Tinker2017a}
Tinker J.~L.,  2017, \mn@doi [Mon. Not. Roy. Astron. Soc.]
  {10.1093/mnras/stx287}, 467, 3533–3541

\bibitem[\protect\citeauthoryear{Tinker, Leauthaud, Bundy, George, Behroozi,
  Massey, Rhodes  \& Wechsler}{Tinker et~al.}{2013}]{Tinker2013}
Tinker J.~L.,  Leauthaud A.,  Bundy K.,  George M.~R.,  Behroozi P.,  Massey
  R.,  Rhodes J.,   Wechsler R.~H.,  2013, \mn@doi [\apj]
  {10.1088/0004-637X/778/2/93}, 778, 93

\bibitem[\protect\citeauthoryear{Tomczak et~al.,}{Tomczak
  et~al.}{2014}]{Tomczak2014}
Tomczak A.~R.,  et~al., 2014, \mn@doi [\apj] {10.1088/0004-637X/783/2/85}, 783,
  85

\bibitem[\protect\citeauthoryear{Wechsler, Bullock, Primack, Kravtsov  \&
  Dekel}{Wechsler et~al.}{2002}]{Wechsler2002}
Wechsler R.~H.,  Bullock J.~S.,  Primack J.~R.,  Kravtsov A.~V.,   Dekel A.,
  2002, \mn@doi [\apj] {10.1086/338765}, 568, 52–70

\bibitem[\protect\citeauthoryear{Weinberg, Mortonson, Eisenstein, Hirata, Riess
   \& Rozo}{Weinberg et~al.}{2013}]{Weinberg2013}
Weinberg D.~H.,  Mortonson M.~J.,  Eisenstein D.~J.,  Hirata C.,  Riess A.~G.,
   Rozo E.,  2013, \mn@doi [\physrep] {10.1016/j.physrep.2013.05.001}, 530,
  87–255

\bibitem[\protect\citeauthoryear{White \& Rees}{White \&
  Rees}{1978}]{White1978}
White S. D.~M.,  Rees M.~J.,  1978, \mn@doi [\mnras] {10.1093/mnras/183.3.341},
  183, 341–358

\bibitem[\protect\citeauthoryear{White, Efstathiou  \& Frenk}{White
  et~al.}{1993}]{White1993}
White S. D.~M.,  Efstathiou G.,   Frenk C.~S.,  1993, \mn@doi [\mnras]
  {10.1093/mnras/262.4.1023}, 262, 1023–1028

\bibitem[\protect\citeauthoryear{Wojtak et~al.,}{Wojtak
  et~al.}{2018}]{Wojtak2018}
Wojtak R.,  et~al., 2018, \mn@doi [\mnras] {10.1093/mnras/sty2257}, 481,
  324–340

\bibitem[\protect\citeauthoryear{Yang, Mo  \& van~den Bosch}{Yang
  et~al.}{2003}]{Yang2003}
Yang X.,  Mo H.~J.,   van~den Bosch F.~C.,  2003, \mn@doi [\mnras]
  {10.1046/j.1365-8711.2003.06254.x}, 339, 1057–1080

\bibitem[\protect\citeauthoryear{Yang, Mo  \& van~den Bosch}{Yang
  et~al.}{2009}]{Yang2009a}
Yang X.,  Mo H.~J.,   van~den Bosch F.~C.,  2009, \mn@doi [\apj]
  {10.1088/0004-637X/693/1/830}, 693, 830–838

\bibitem[\protect\citeauthoryear{Zentner, Hearin  \& van~den Bosch}{Zentner
  et~al.}{2014}]{Zentner2014}
Zentner A.~R.,  Hearin A.~P.,   van~den Bosch F.~C.,  2014, \mn@doi [\mnras]
  {10.1093/mnras/stu1383}, 443, 3044–3067

\bibitem[\protect\citeauthoryear{Ziparo et~al.,}{Ziparo
  et~al.}{2016}]{Ziparo2016}
Ziparo F.,  et~al., 2016, \mn@doi [\aap] {10.1051/0004-6361/201526792}, 592, A9

\bibitem[\protect\citeauthoryear{Zu \& Mandelbaum}{Zu \&
  Mandelbaum}{2015}]{Zu2015}
Zu Y.,  Mandelbaum R.,  2015, \mn@doi [\mnras] {10.1093/mnras/stv2062}, 454,
  1161–1191

\makeatother
\end{thebibliography}
\label{lastpage}

\end{document}